\documentclass[twocolumn,a4paper,accepted=2019-07-03]{quantumarticle}
\pdfoutput=1
\usepackage{graphicx}
\usepackage[pdftex,colorlinks=true,linkcolor=blue,citecolor=blue,urlcolor=blue]{hyperref}
\usepackage{amsmath, amsthm, amssymb}
\usepackage{subfigure}
\usepackage{comment}
\usepackage{url}
\usepackage{tikz}
\usepackage{ragged2e}
\usepackage[numbers,sort&compress]{natbib}

\newcommand{\R}{\mathbb{R}}
\newcommand{\N}{\mathbb{N}}

\newcommand{\braket}[1]{\langle #1 \rangle}

\newcommand{\proj}[1]{| #1 \rangle \langle #1 |}

\DeclareMathOperator{\poly}{poly}

\newcommand{\be}{\begin{equation}}
\newcommand{\ee}{\end{equation}}
\newcommand{\bea}{\begin{eqnarray}}
\newcommand{\eea}{\end{eqnarray}}
\newcommand{\bes}{\begin{equation*}}
\newcommand{\ees}{\end{equation*}}
\newcommand{\beas}{\begin{eqnarray*}}
\newcommand{\eeas}{\end{eqnarray*}}

\makeatletter
\newtheorem*{rep@theorem}{\rep@title}
\newcommand{\newreptheorem}[2]{%
\newenvironment{rep#1}[1]{%
 \def\rep@title{#2 \ref{##1} (restated)}%
 \begin{rep@theorem}}%
 {\end{rep@theorem}}}
\makeatother

\newcommand{\boxalgm}[3]{
\renewcommand{\figurename}{Algorithm}
\begin{figure}[tb]
\noindent \framebox{
\begin{minipage}{.45\textwidth}
\justify
#3
\end{minipage}
}
\caption{#2}
\label{#1}
\end{figure}
\renewcommand{\figurename}{Figure}
}

\newtheorem{thm}{Theorem}
\newtheorem*{thm*}{Theorem}

\newtheorem*{lem*}{Lemma}
\newtheorem{claim}[thm]{Claim}

\newreptheorem{thm}{Theorem}
\newreptheorem{lem}{Lemma}
\newreptheorem{cor}{Corollary}

\input{Qcircuit.tex}

\begin{document}

\title{Applying quantum algorithms to constraint satisfaction \mbox{problems}}

\author{Earl Campbell}
\affiliation{Department of Physics and Astronomy, University of Sheffield, Sheffield, UK}

\author{Ankur Khurana}
\affiliation{Quantum Engineering Centre for Doctoral Training, University of Bristol, UK}
\affiliation{School of Physics, University of Bristol, UK}

\author{Ashley Montanaro}
\affiliation{School of Mathematics, University of Bristol, UK}
\email{ashley.montanaro@bristol.ac.uk}

\begin{abstract}
Quantum algorithms can deliver asymptotic speedups over their classical counterparts. However, there are few cases where a substantial quantum speedup has been worked out in detail for reasonably-sized problems, when compared with the best classical algorithms and taking into account realistic hardware parameters and overheads for fault-tolerance. All known examples of such speedups correspond to problems related to simulation of quantum systems and cryptography. Here we apply general-purpose quantum algorithms for solving constraint satisfaction problems to two families of prototypical NP-complete problems: boolean satisfiability and graph colouring. We consider two quantum approaches: Grover's algorithm and a quantum algorithm for accelerating backtracking algorithms. We compare the performance of optimised versions of these algorithms, when applied to random problem instances, against leading classical algorithms. Even when considering only problem instances that can be solved within one day, we find that there are potentially large quantum speedups available. In the most optimistic parameter regime we consider, this could be a factor of over $10^5$ relative to a classical desktop computer; in the least optimistic regime, the speedup is reduced to a factor of over $10^3$. However, the number of physical qubits used is extremely large, and improved fault-tolerance methods will likely be needed to make these results practical. In particular, the quantum advantage disappears if one includes the cost of the classical processing power required to perform decoding of the surface code using current techniques.
\end{abstract}

\maketitle
\setlength{\parskip}{3pt}



\noindent Many quantum algorithms are known, for tasks as diverse as integer factorisation~\cite{shor94} and computing Jones polynomials~\cite{aharonov06}. Indeed, at the time of writing, the Quantum Algorithm Zoo website~\cite{qazoo} cites 392 papers on quantum algorithms. However, there are relatively few cases known where quantum algorithms substantially outperform their classical counterparts for problems of practical importance, and the runtime of the quantum algorithm has been calculated in detail. Examples include simulating the chemical processes involved in biological nitrogen fixation~\cite{reiher17}; breaking cryptosystems based on integer factorisation~\cite{vanmeter06,haner17} and elliptic curves~\cite{roetteler17}; quantum simulation of spin systems~\cite{childs17} and electronic structure Hamiltonians~\cite{babbush18}. In all of these cases, the underlying quantum algorithm achieves an exponential speedup over its classical counterpart, and under realistic assumptions about the performance of the quantum hardware, can solve a problem in days or weeks that might take decades or centuries on a fast classical computer.

Notwithstanding the extreme practical importance of some of these applications, they share the feature that they are rather special-purpose. While simulation of quantum systems, for example, has a large number of uses~\cite{georgescu14}, there are many problem domains for which it is simply not relevant. Here we focus on problems in the general area of constraint satisfaction and optimisation -- an area critical to many different industry sectors and applications -- and aim to quantify the likely advantage that could be achieved by quantum computers. We seek to satisfy the following desiderata:
\begin{enumerate}
\item {\bf (Rigour)} There should exist a quantum algorithm which solves the problem with provable correctness and rigorous performance bounds.

\item {\bf (Broad utility)} The abstract problem solved by the algorithm should be broadly useful across many different applications.

\item {\bf (Performance bounds)} We should compute the performance of the quantum and classical algorithms explicitly for particular problem instances, including all relevant overheads.

\item {\bf (Runtime)} The problem instance used for comparison should be one that can be solved by the quantum computer within a reasonable time (e.g.~$<1$ day) under reasonable assumptions about the performance of the quantum hardware.

\item {\bf (Benchmarking)} The point of comparison should be one of the best classical algorithms known, running on modern-day hardware.
\end{enumerate}

These points between them seem to put severe restrictions on the ability of quantum computing to achieve a significant performance enhancement. First, the requirement of rigour rules out heuristic algorithms running on current or near-term hardware, such as quantum annealing (e.g.~\cite{ronnow14}) or the quantum approximate optimisation algorithm~\cite{farhi14}.

Next, the requirement of broad utility rules out the exponential speedups discussed above. In general, quantum algorithms that are broadly applicable to accelerating classical algorithms tend to achieve at best quadratic speedups (that is, the scaling with problem size of the quantum algorithm's runtime is approximately the square root of its classical counterpart); one famous example is Grover's algorithm~\cite{grover97}, which speeds up unstructured search. In other models, such as query complexity, it can even be proven that special problem structure is required to see an exponential quantum speedup~\cite{beals01}. Although even a quadratic speedup will become arbitrarily large for large enough input sizes, choosing an extremely large input size will make the execution time of the quantum algorithm unacceptably long for practical purposes. This motivates the runtime requirement, which is particularly challenging to satisfy because many quantum algorithms (e.g.\ Grover's algorithm~\cite{zalka99}) are inherently serial: they cannot be parallelised without reducing the quantum speedup.

The requirement to compute accurate performance bounds implies that we should take into account not just the performance of the quantum hardware itself (which will in general be slower than modern-day classical hardware) but also the overhead from fault-tolerance, which could correspond to an increase of several orders of magnitude in the number of qubits required, and a concomitant increase in cost and runtime. Table \ref{tab:params} lists parameters for quantum hardware in various regimes (``Realistic'', ``Plausible'' and ``Optimistic''). ``Realistic'' is based on relatively modest improvements on parameters already demonstrated in experiment. For example, in superconducting qubit systems, 2-qubit gate times of 40ns~\cite{barends14} and measurement times of under 50ns~\cite{walter17} have been demonstrated, and numerical simulations suggest the possibility of 26ns gates~\cite{deng17}; 2-qubit gate error rates of 0.06 have been demonstrated~\cite{barends14} and 0.004 is predicted~\cite{deng17}. In ion traps, 2-qubit gate times of 480ns have been demonstrated~\cite{schafer18}, as have error rates of 0.001 (at the cost of increasing the gate duration)~\cite{ballance16}. The other categories are based on the simple assumption that order-of-magnitude improvements are possible.

One may reasonably query whether this assumption is realistic. Considering gate times, there is quite a wide variation in leading results reported in the literature. Even considering superconducting qubits alone, these include 40ns in a 5-qubit system (2014)~\cite{barends14}, 150ns in a 6-qubit system (2017)~\cite{kandala17}, and 100--250ns in a 19-qubit system (2017)~\cite{otterbach17}. Classically, within the period 1995-2000 Intel CPUs increased in clock speed by about a factor of 10. In the case of error rates, although error rates of $10^{-3}$ combined with $<$100ns gate times have not yet been demonstrated, an ultimate error rate of $10^{-5}$ may even be pessimistic, if more exotic technologies such as topological quantum computers come to fruition; an effective error rate of $10^{-9}$ has been assumed elsewhere in the literature for such devices~\cite{reiher17}. See~\cite{aggarwal17} for a more detailed performance extrapolation.

\begin{table}[t]
\fontfamily{cmss}\selectfont
\begin{center}
\resizebox{0.48\textwidth}{!}{
\begin{tabular}{|c|c|c|c|}
\hline Parameter & Realistic & Plausible & Optimistic \\
\hline Measurement time & 50ns & 5ns & 0.5ns \\
2-qubit gate time & 30ns & 3ns & 0.3ns \\
Cycle time & 200ns & 20ns & 2ns \\
Gate error rate & $10^{-3}$ & $10^{-4}$ & $10^{-5}$ \\
\hline
\end{tabular}
}
\end{center}
\caption{Parameter regimes considered in this work. ``Realistic'' corresponds to values reported in the literature as possible now, or predicted in the near future. ``Cycle time'' is the time required to perform one surface code cycle. Each such cycle comprises four 2-qubit gates, possibly two 1-qubit gates and a measurement. These must be performed for each of X and Z, but this can be parallelised to an extent that depends on the relative times required to implement measurements and gates; we therefore only consider one X/Z cycle when estimating the cycle time, and assume that a 1-qubit gate can be implemented in half the time required for a 2-qubit gate.}
\label{tab:params}
\end{table}

Finally, the benchmarking requirement implies that we should not simply compare the quantum algorithm against the simplest or most obvious classical competitor, but should choose a competitor that is one of the fastest options actually used in practice. For example, Grover's algorithm can determine satisfiability of a boolean formula containing $n$ variables with $O(2^{n/2})$ evaluations of the formula~\cite{grover97}, whereas exhaustive search would use $O(2^n)$ evaluations. However, other algorithms are known which are much faster in practice, for example based on the DPLL method~\cite{davis62,davis60}. A fair quantum-classical comparison should test against these best algorithms.

It is worth pausing to check whether there is hope to achieve a substantial quantum speedup at all while satisfying all the above desiderata. Imagine there exists a family of problems which is exceptionally challenging to solve classically: for a problem instance involving $n$ boolean variables, the best classical algorithm consists of simply evaluating some simple ``oracle'' function of the variables for $\sim 2^n$ different assignments. Further assume that this function can be evaluated efficiently on both a classical and quantum computer. For example, we could consider a cryptographic hash function. Such functions are designed to be easy to compute and hard to invert, and in some cases the hash of (e.g.)\ a 256-bit integer can be computed in under approximately 1000 CPU cycles~\cite{ebacs}. Given the overhead required to implement a classical circuit reversibly, it is hard to imagine\footnote{However, see Section \ref{sec:grover} for a very low-depth quantum circuit for boolean satisfiability.} performing an equivalently complex operation via a quantum circuit in circuit depth less than 1000 (and if this were possible, it is plausible that it would lead to a faster classical algorithm).

Therefore, assume that the quantum circuit depth required to solve an instance of size $n$ is approximately $1000 \times 2^{n/2}$,  corresponding to approximately the depth required to execute Grover's algorithm, while the classical runtime is $1000 \times 2^n$ clock cycles. For simplicity, assume the classical computer's clock speed is 1GHz. (This may appear unrealistic, as high-performance computing hardware could be used to solve a problem of this form via parallel computation. However, in the context of a cost comparison between quantum and classical computation, this would correspond to multiplying the cost of the classical computation by the number of parallel processors used. So computing the speedup over one classical processor can be used as a proxy for the cost advantage over multiple processors.)

Given no overhead at all for fault-tolerance, considering the gate times in Table \ref{tab:params} and only problem instances that can be solved in 1 day, we obtain the middle row of Table \ref{tab:noft}. It is clear that the speedups achieved are very substantial in all cases. An example of a more realistic depth overhead is the quantum circuit for computing the SHA-256 hash function described in~\cite{amy16}, which has depth $\approx 5 \times 10^5$. Using this example, we achieve a speedup factor between roughly $2 \times 10^2$ and $4 \times 10^6$, depending on assumptions, which is still quite substantial at the high end. Note that, counterintuitively, decreasing the quantum clock speed (equivalently, increasing the oracle circuit depth) by a factor of $c$ reduces the largest speedup that can be achieved in a given time period by a factor of approximately $c^2$. This strongly motivates the design of depth-efficient quantum circuits and hardware with high clock speeds.

Table \ref{tab:noft} represents an estimate of the best possible speedups for square-root-type quantum algorithms. It remains to attempt to show that significant speedups can actually be achieved for problems of practical interest, which is our focus in this work.

\begin{table}[t]
\fontfamily{cmss}\selectfont
\resizebox{0.48\textwidth}{!}{
\begin{tabular}{|c|c|c|c|c|}
\hline Oracle & & Realistic & Plausible & Optimistic \\
depth & & & &\\
\hline & Max depth & $2.88 \times 10^{12}$ & $2.88 \times 10^{13}$ & $2.88 \times 10^{14}$ \\
\hline 1000 & Max size $n$ & 62 & 69 & 76 \\
& Cl.\ runtime & $4.61 \times 10^{12}s$ & $5.90 \times 10^{14}s$ & $7.56 \times 10^{16}s$ \\
& Speedup & $7.16 \times 10^7$ & $8.10 \times 10^{9}$ & $9.16 \times 10^{11}$ \\
\hline $5 \times 10^5$ & Max size $n$ & 44 & 51 & 58 \\
\cite{amy16} & Cl.\ runtime & $1.76 \times 10^7 s$ & $2.25 \times 10^9s$ & $2.88 \times 10^{11}s$ \\
& Speedup & $2.80 \times 10^2$ & $3.16 \times 10^4$ & $3.58 \times 10^6$ \\
\hline
\end{tabular}
}
\caption{Likely upper bounds on speedup factors possible for square-root-type quantum algorithms running for at most one day in different regimes, assuming that there is no overhead for fault tolerance, so maximum circuit depths are only determined by gate times.}
\label{tab:noft}
\end{table}


\section{Our results}
\label{sec:results}

In an attempt to satisfy all the above requirements, we focus on two prominent and fundamental NP-complete problems: graph colouring and boolean satisfiability. In the graph colouring problem, we are given a graph $G$ with $n$ vertices, and asked to assign one of $k$ colours to each vertex, such that no pair of adjacent vertices shares the same colour. If no such colouring exists, we should detect this fact. In the boolean satisfiability problem, we are given a boolean formula $\phi$ on $n$ variables in conjunctive normal form and asked to find a satisfying assignment to the formula, if one exists. That is, the formula is made up of clauses, where each clause is an OR function of some of the variables (each possibly appearing negated), and we are asked to find an assignment to the variables such that all of the clauses evaluate to true. Here we consider the special case $k$-SAT, where each clause contains exactly $k$ variables.

Each of these problems has countless direct applications. In the case of graph colouring, these include register allocation~\cite{chow90}; scheduling~\cite{leighton79}; frequency assignment problems~\cite{aardal07}; and many other problems in wireless networking~\cite{balasundaram06}. In the case of boolean satisfiability, these include formal verification of electronic circuits~\cite{prasad05}; planning~\cite{selman92}; and computer-aided mathematical proofs~\cite{konev15}.

We seek a problem instance which can be solved using a quantum computer in one day, but would take substantially longer for a classical computer to solve. This raises the question of how to be confident that the runtime of the classical algorithm is indeed large (we cannot simply run the algorithm, as by definition it would take too long). A strategy to achieve this is to find a family of instances, parametrised by problem size, which can be solved for small problem sizes in a reasonable time, and where the runtime for larger problem sizes can be extrapolated from these.

A straightforward way to satisfy this criterion is to choose instances at random. Another advantage of using random instances is that they are likely to be hard for classical algorithms, as they have no structure that the algorithm can exploit. Indeed, in the case of graph colouring, even random instances on around 80 vertices are already challenging for the best classical algorithms~\cite{malaguti10}. We use the following models:

\begin{itemize}
\item $k$-colouring: pick a uniformly random (Erd\H{o}s-R\'enyi) graph on $n$ vertices, where each edge is present with probability 0.5. As $n \rightarrow \infty$, the chromatic number $\chi_{n,0.5}$ of such graphs has long been known to be $(1+o(1)) n / (2 \log_2 n)$ with high probability~\cite{bollobas88}. Empirically, the estimate
\be \label{eq:chromest} \chi_{n,0.5} \approx \frac{n}{2 \log_2 n - 2\log_2 \log_2 n - 1}, \ee
which is based on a small modification to the asymptotic formula in~\cite{panagiotou09}, seems to be an excellent predictor of the mean chromatic number of a random graph (see Figure \ref{fig:chrom}). For $n \le 200$, this estimate is at most 24.

\begin{figure}
\begin{center}
\includegraphics[width=0.5\textwidth]{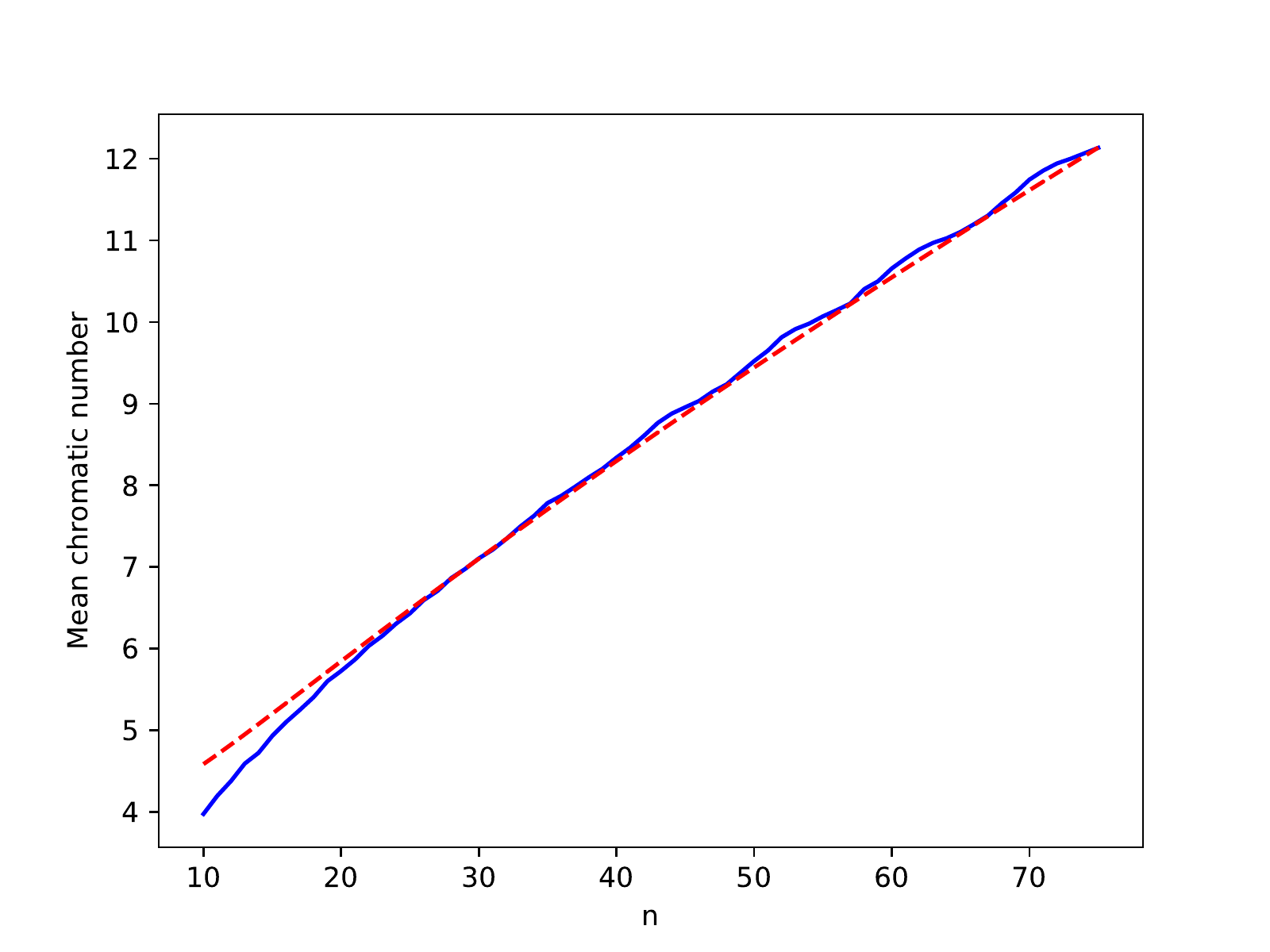}
\end{center}
\caption{Blue solid line: the mean chromatic number of 1000 random graphs on $n$ vertices with edge probability $1/2$. Red dashed line: estimate from (\ref{eq:chromest}). For $n \ge 20$, all 1000 graphs had chromatic number within $\pm2$ of the estimate.}
\label{fig:chrom}
\end{figure}

\item $k$-SAT: choose $m$ clauses, each of which contains $k$ variables. Each clause is picked independently and uniformly at random from the set of $2^k \binom{n}{k}$ distinct clauses containing $k$ distinct variables. We aim to fix $m$ such that $m / n \approx \alpha_k$, where $\alpha_k$ is the threshold for $k$-SAT. The threshold is the point $\alpha_k$ such that, as $n \rightarrow \infty$, a random $k$-SAT formula on $\alpha n$ clauses will be satisfiable with probability approaching 1 for $\alpha < \alpha_k$, and unsatisfiable with probability approaching 1 for $\alpha > \alpha_k$. It has long been predicted theoretically, and verified experimentally, that random $k$-SAT instances around the satisfiability threshold will be very challenging~\cite{cheeseman91}.
\end{itemize}

Here we applied efficient quantum algorithms with rigorous performance bounds to these random instances of $k$-colouring and $k$-SAT problems, carried out a detailed analysis of their performance (including overheads for fault-tolerance), and compared them against leading classical competitors.

It may be debatable whether random problem instances satisfy the broad utility criterion above, as they may not correspond to instances encountered in practice. Indeed, SAT solvers are often able to solve significantly larger instances of structured SAT problems than random problems. However, in this work we aim to find problem families on which quantum algorithms achieve as large a speedup as possible, which is approximately equivalent to finding the hardest problem instances possible for the classical algorithm. Removing structure is one way to achieve this, while remaining within a family of problems that do contain many practically relevant instances. Random problem instances could also be seen as modelling the unstructured and most challenging component of a partially structured problem.

We considered two (families of) general-purpose quantum algorithms: Grover's algorithm~\cite{grover97} for accelerating unstructured search, and a quantum algorithm for accelerating the general classical algorithmic technique known as backtracking~\cite{montanaro18}. Each of these algorithms achieves a near-quadratic reduction in computational complexity compared with its classical counterpart (that is, if the classical runtime is $T$, the dominant component of the quantum runtime scales like $\sqrt{T}$) and has a rigorous correctness proof.

In the case of $k$-SAT, we compared the performance of these two algorithms against the performance of the Maple\_LCM\_Dist SAT solver, which was the winner of the SAT Competition 2017\footnote{A modified version of this solver was also the winner of the 2018 competition.}~\cite{balyo17}. We evaluated the performance of this solver on many random instances, for different values of $k$, to estimate its runtime scaling with the number of variables $n$. We then calculated the complexity of highly optimised versions of Grover's algorithm and the quantum backtracking algorithm applied to this problem. In order to solve the largest instances possible while meeting the runtime requirement, the algorithms are optimised to perform as many operations in parallel as possible, and hence minimise their circuit depths.

In the case of graph $k$-colouring, we compared against the commonly used (``de facto standard''~\cite{sewell96}) DSATUR algorithm~\cite{brelaz79} (see Section \ref{sec:dsatur}). This is a backtracking algorithm itself, so can be accelerated directly via the quantum backtracking algorithm. In this case, Grover's algorithm is not applicable, as for relevant values of $k$ the runtime of DSATUR is empirically exponentially faster than the $O(k^{n/2})$ runtime scaling that would be achieved by Grover's algorithm applied to $k$-colouring.

\begin{figure}
\includegraphics[width=\columnwidth]{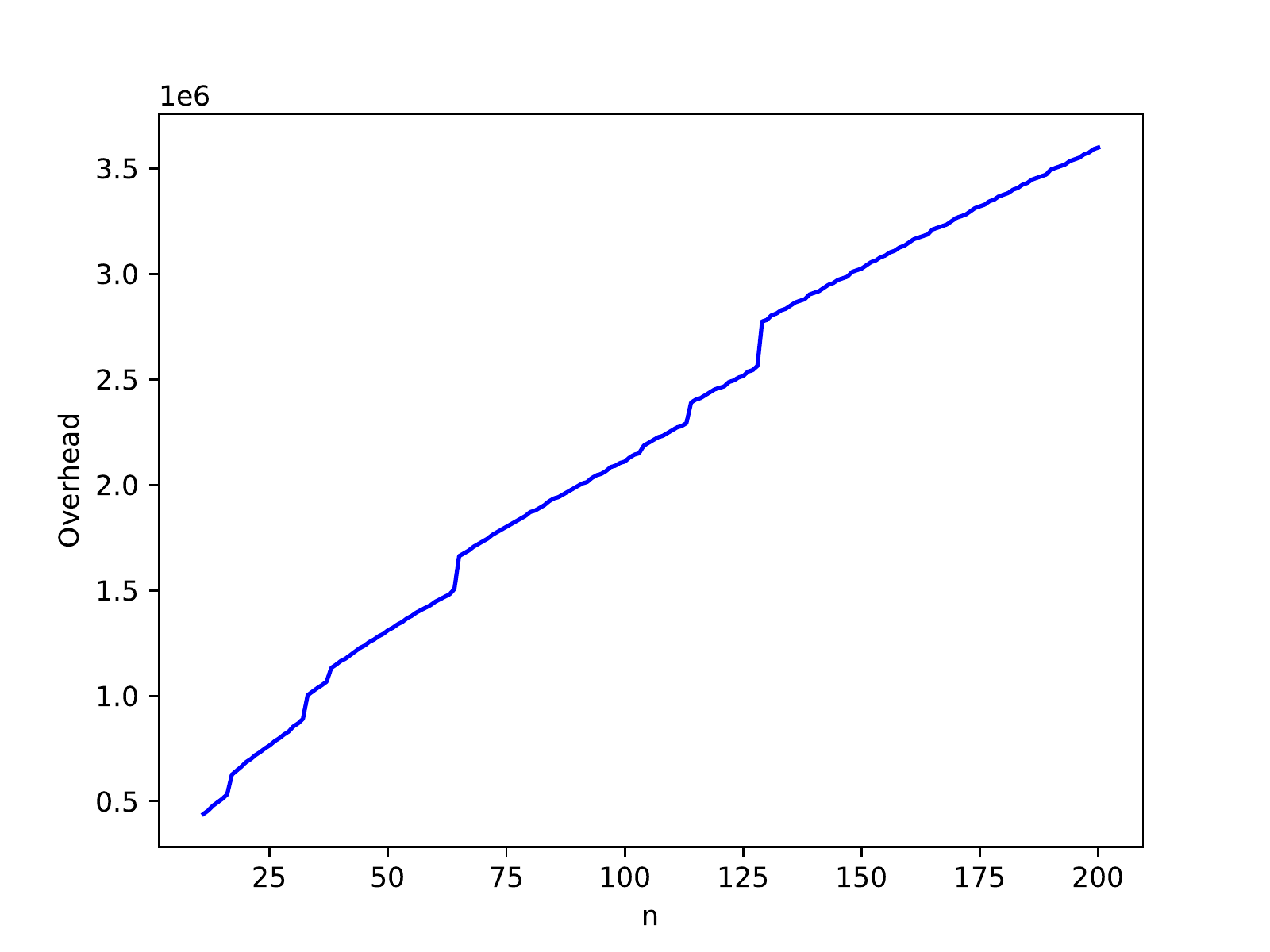}
\caption{The runtime (circuit depth) of the quantum algorithm for backtracking is of the form $f(n,k) \sqrt{T}$, where $T$ is the number of nodes in the backtracking tree. Figure illustrates scaling of $f(n,k)$ with $n$ when $k$ is chosen according to (\ref{eq:chromest}).}
\label{fig:overhead}
\end{figure}

In Figure \ref{fig:overhead} we illustrate the depth overhead of our optimised $k$-colouring algorithm versus the input size $n$. For reasonable graph sizes (e.g.\ $n \in \{100,\dots,200\}$) it is less than $4 \times 10^6$, and hence not substantially greater than the overhead of the SHA-256 hash function implemented in~\cite{amy16}. We stress, however, that the algorithm has been optimised for depth, and the number of logical qubits that it uses is large ($\gg 10^5$ for reasonable problem sizes).

We obtain the result that for both $k$-SAT and $k$-colouring, substantial quantum speedups could be possible: in the case of $k$-SAT, in the most optimistic regime we consider, the speedup could be as large as a factor of over $10^5$, compared with a standard desktop computer. (That is, to solve an instance solved by the quantum algorithm in one day, the classical algorithm running on a standard computer would need over $10^5$ days.) In the case of $k$-colouring, speedups by a factor of over $10^4$ could be possible. However, the extent of these speedups is strongly dependent on the details of the hardware parameters, and the overhead for error-correction. In other regimes for these, there is no quantum speedup at all. In addition, the number of physical qubits required to obtain these speedups is very large (e.g.\ over $10^{12}$). This is largely caused by the need for many large ``factories'' to operate in parallel to produce high-quality magic states, which are used to implement T gates and Toffoli gates fault-tolerantly in the error-correcting code used (the surface code~\cite{fowler12}). A related issue is that this speedup does not take into account the cost of classical processing in the quantum error-correction process, which should also be considered to obtain a true cost comparison (see Section \ref{sec:classcost}). When we include an estimate for the cost of the classical processing power required to perform decoding of the surface code using current techniques, the quantum advantage disappears. Thus, improvements to fault-tolerance methods are likely to be required if such speedups are to be realised in practice.

In order to state our results more precisely, we must describe the model and methodology used to calculate the cost of a quantum computation.


\subsection{Timing and cost model}
\label{sec:model}

Here we outline our resource methdology, which follows a model developed by several previous works in this area~\cite{fowler12,amy16,ogorman17,aggarwal17,babbush18}. The model assumes that the quantum computation is encoded using the surface code~\cite{fowler12}, a quantum error-correcting code with properties that make it an excellent candidate for implementation on near-term hardware platforms (e.g.\ high fault-tolerance threshold, implementability via local operations). Then the cost of a computation can be calculated via the following sequence of steps:

\begin{enumerate}
\item Determine the cost of the quantum circuit, in terms of the number of gates used and the circuit depth.

\item Calculate the number of physical qubits required for the logical computation, and the physical depth.

\item Insert hardware-dependent parameters for clock speed and other variables to compute a physical runtime. According to the runtime requirement, this should be at most 1 day, putting a limitation on the problem instance size that can be solved.

\item Use the above to make a comparison between the cost of quantum and classical computation.
\end{enumerate}

When considering the cost of quantum circuits implemented using the surface code, it is helpful to divide the circuit into parts consisting of Clifford gates (which can be implemented relatively straightforwardly) and non-Clifford gates (which cannot). In the circuits that we consider, the non-Clifford gates used are Toffoli and T gates.

\begin{table}[t]
\begin{center}
\begin{tabular}{|c|c|c|c|}
\hline $N$ $\backslash$ $\epsilon$ & $10^{-3}$ & $10^{-4}$ & $10^{-5}$ \\
\hline $10^{12}$ & $4.10 \times 10^{7}$ & $4.22 \times 10^{6}$ & $8.98 \times 10^{5}$\\
\hline $10^{18}$ & $2.45 \times 10^{8}$ & $9.86 \times 10^{6}$ & $2.30 \times 10^{6}$\\
\hline $10^{24}$ & $4.51 \times 10^{8}$ & $4.60 \times 10^{7}$ & $4.69 \times 10^{6}$\\
\hline
\end{tabular}
\end{center}
\caption{Representative spacetime costs (measured in units of surface code cycles $\times$ physical qubits) to implement one Toffoli gate, assuming gate error rate $\epsilon$ and a circuit of $N$ Toffoli gates. Calculated using method described in Appendix \ref{app:tofffactory}. $10^{24}$ gates is a generous upper bound on the number of Toffoli gates that can be executed in 1 day (corresponding to $>10^9$ qubits at a clock speed of 1GHz).}
\label{tab:factory}
\end{table}

Toffoli and T gates can be implemented fault-tolerantly using a state injection technique where a special state is prepared offline (a Toffoli state~\cite{jones13,eastin13} or T state~\cite{bravyi05a}),  and then used to implement the corresponding gate. We include in Appendix \ref{app:tofffactory} an algorithm for computing the costs associated with this, based on the protocol of~\cite{eastin13,jones13} and using the analysis of~\cite{ogorman17} (see also~\cite{fowler12,aggarwal17}). Some illustrative spacetime costs are shown in Table \ref{tab:factory} for Toffoli gates, which dominate the complexity of the circuits we consider; the values for T gates are similar.

For reasonable parameter ranges for the error rate $\epsilon$ and the number $N$ of Toffoli gates, and using standard protocols, the number of qubits used by a single Toffoli-factory is between $10^4$ and $10^6$, and the depth of the factory is between 100 and 1000 surface code cycles. However, using more factories this process can be parallelised, such that each new magic state is available almost arbitrarily quickly. Using time-optimal methods~\cite{fowler12a}, the limiting factor becomes only the time required to inject a magic state, which is the time taken for a single physical measurement\footnote{There is also a cost associated with performing a gate before the measurement, but when multiple logical gates are performed, this cost becomes negligible.}.

The time complexity of the circuit is then governed by its depth, considering only Toffoli or T gates. As a Toffoli gate can be implemented using a single layer of T gates~\cite{selinger13} or injected directly from a Toffoli magic state, this is equal to the ``T-depth'' of the circuit. The T-depth is defined as the number of T-stages in a circuit, where a T-stage is a group of T gates that can be performed simultaneously~\cite{amy13}. Each time step corresponds to the cost of one measurement.

The parts of the circuit corresponding to Clifford gates can also be implemented using state injection by preparing a particular graph state offline, then measuring all the qubits of this state simultaneously. As the results of this measurement only affect the interpretation of subsequent measurement results, not which measurements are performed, it can be performed in parallel with the implementation of a subsequent Toffoli or T gate. Hence Clifford gates do not contribute to the time cost of the circuit.

The drawback of implementing the circuit in this way is that a large number of ancillas are used, though in practice this ancilla cost is still small compared to the size of the magic state factory.  Making a detailed analysis of time-optimal implementations of a Grover oracle, we found that the factory comprised $95\%-99\%$ of all physical qubits, so it is safe to assume factory-dominated costs. There is a space-time tradeoff here and we have chosen to minimise time over space.

Some additional aspects which we do not take into account in our cost calculations, for simplicity and because of their hardware-specific nature, are:

\begin{itemize}
\item Any additional cost required to implement long-range gates. This cost will depend on the underlying hardware (e.g.\ certain architectures allow long-range gates, while others are restricted to nearest-neighbour interactions), and some apparently ``long-range'' gates can be implemented efficiently in the surface code (e.g.\ controlled-NOT gates with multiple targets).

\item Any additional cost required to lay out and route qubits physically within a desired spacetime volume. A discussion of these issues can be found in~\cite{babbush18}.
\end{itemize}

One way to address point 4 above, and find a basis for comparing the cost of classical and quantum computation, is to consider the cost of the classical processing required to perform the quantum computation (and in particular to carry out the calculations required for fault-tolerance). We discuss this in Section \ref{sec:classcost}.


\subsection{Summary of results}
\label{sec:summary}

\begin{table}[t]
\fontfamily{cmss}\selectfont
\resizebox{0.48\textwidth}{!}{
\begin{tabular}{|c|c|c|c|}
\hline & Realistic & Plausible & Optimistic \\
\hline Max $n$ & 65 & 72 & 78 \\
T-depth & $1.46 \times 10^{12}$ & $1.65 \times 10^{13}$ & $1.32 \times 10^{14}$\\
Toffoli count & $4.41 \times 10^{17}$ & $5.52 \times 10^{18}$ & $4.79 \times 10^{19}$ \\
Factory qubits & $3.14 \times 10^{13}$ & $5.15 \times 10^{12}$ & $1.38 \times 10^{12}$ \\
Speedup factor & $1.62 \times 10^{3}$ & $1.73 \times 10^{4}$ & $1.83 \times 10^{5}$ \\
\hline
\end{tabular}
}
\caption{Likely speedup factors for 14-SAT via Grover's algorithm achievable in different regimes.}
\label{tab:speedups_satgrover}
\end{table}

\begin{table}[t]
\fontfamily{cmss}\selectfont
\resizebox{0.48\textwidth}{!}{
\begin{tabular}{|c|c|c|c|}
\hline & Realistic & Plausible & Optimistic \\
\hline Max $n$ & 55 & 63 & 72 \\
T-depth & $1.63 \times 10^{12}$ & $1.43 \times 10^{13}$ & $1.63 \times 10^{14}$\\
T/Toffoli count & $4.72 \times 10^{18}$ & $4.72 \times 10^{19}$ & $6.16 \times 10^{20}$ \\
Factory qubits & $3.85 \times 10^{14}$ & $5.03 \times 10^{13}$ & $2.17 \times 10^{13}$ \\
Speedup factor & $1.50 \times 10^{1}$ & $3.92 \times 10^{2}$ & $1.16 \times 10^{4}$ \\
\hline
\end{tabular}
}
\caption{Likely speedup factors for 12-SAT via backtracking achievable in different regimes.}
\label{tab:speedups_satbt}
\end{table}

\begin{table}[t]
\fontfamily{cmss}\selectfont
\resizebox{0.48\textwidth}{!}{
\begin{tabular}{|c|c|c|c|}
\hline & Realistic & Plausible & Optimistic \\
\hline Max $n$ & 113 & 128 & 144 \\
T-depth & $1.70 \times 10^{12}$ & $1.53 \times 10^{13}$ & $1.62 \times 10^{14}$\\
T/Toffoli count & $8.24 \times 10^{17}$ & $9.94 \times 10^{18}$ & $1.24 \times 10^{20}$ \\
Factory qubits & $6.29 \times 10^{13}$ & $9.26 \times 10^{12}$ & $3.59 \times 10^{12}$ \\
Speedup factor & $7.25 \times 10^{0}$ & $5.17 \times 10^{2}$ & $4.16 \times 10^{4}$ \\
\hline
\end{tabular}
}
\caption{Likely speedup factors for graph colouring via backtracking achievable in different regimes.}
\label{tab:speedups}
\end{table}

Having described the cost model, we summarise the results obtained in Tables \ref{tab:speedups_satgrover} to \ref{tab:speedups}. Each table column corresponds to an extrapolation for the maximal instance size $n$ that can be solved by a quantum algorithm in one day, and includes the parameters of this algorithm, and the speedup obtained. These speedups are expressed as a multiple of the likely performance of the DSATUR and Maple\_LCM\_Dist algorithms running on a standard desktop computer in the cases of graph colouring and SAT, respectively (see Section \ref{sec:classexp} for more on the classical experimental results, and the assumptions made). We stress that these figures are sensitive to the precise assumptions made about the classical algorithm's scaling and hardware performance, as well as to certain assumptions (detailed below) about the quantum algorithms' performance on random instances\footnote{The runtime of the quantum algorithm for graph colouring experiences a small overhead that varies depending on the instance (see Section \ref{sec:classexpgc}), and the complexity of a circuit synthesis step used in the algorithm is also problem-dependent~\cite{bocharov15} (see Section \ref{sec:diffusion}).}. However, any reduction in performance due to a change in these assumptions can be offset by allowing the quantum algorithm to run for longer.

If there were no need for fault-tolerance at all, the runtime of the algorithm would be determined only by the time required for 2-qubit gates, which is somewhat faster than the measurement time in Table~\ref{tab:params}, so the speedup factor would likely be somewhat larger.

The results in Tables \ref{tab:speedups_satgrover} to \ref{tab:speedups} were obtained by using computer programs to calculate the complexity of the various algorithms used (in terms of T-depth and T-count) for different parameter values. We then chose parameters that produced the largest speedups, while respecting the constraint that the quantum algorithm should run for at most one day. For example, in the case of $k$-SAT and Grover's algorithm, choosing $k=14$ led to the largest quantum speedup. All code developed, together with the experimental results for the classical algorithms, is available at~\cite{code}.

The largest potential speedup factor found is reasonably large in the ``Plausible'' scenario, and very large in the ``Optimistic'' scenario; over $10^5$ in the case of applying Grover's algorithm to random 14-SAT, and over $4 \times 10^4$ in the case of determining colourability of a random graph with 144 vertices. However, the number of physical qubits used is very large, which (as discussed in Section \ref{sec:classcost}) implies a concomitant increase in the cost of classical processing, which could erase this advantage. This strongly motivates the design of improved fault-tolerance strategies. Observe that this overhead could be mitigated somewhat at the expense of allowing a longer runtime.

It is interesting to note that, in the case of $k$-SAT, the quantum backtracking algorithm achieves worse performance than straightforward use of Grover's algorithm. This is because of lower-order terms in the runtime (cf.\ Tables \ref{tab:complexities_grover} and \ref{tab:complexities_bt_ksat} below); although the backtracking algorithm will be more efficient for large problems, Grover's algorithm is faster for the problem sizes that can be solved in one day.


\subsection{Organisation and notation}

In the remainder of this paper, we give the technical details behind the calculations reported in Tables \ref{tab:speedups_satgrover} to \ref{tab:speedups}. First, in Sections \ref{sec:grover} and \ref{sec:backtrack}, we describe the variants of Grover's algorithm and the backtracking algorithm that we use. In Section \ref{sec:btopt}, we discuss the detailed implementation decisions and optimisations that go into calculating the backtracking algorithm's complexity in the case of graph colouring. Section \ref{sec:btksat} describes the modifications that need to be made to apply the algorithm to $k$-SAT. Section \ref{sec:dsatur} describes the classical DSATUR algorithm, while Section \ref{sec:classexp} gives the results of the classical experiments to determine the empirical complexity of Maple\_LCM\_Dist and DSATUR. Section \ref{sec:classcost} discusses how to estimate the cost of quantum computation in terms of classical processing. We finish in Section \ref{sec:conclusions} with conclusions and further discussion.

We use certain notation throughout the paper. All logs are base 2 and $[n]$ denotes the set $\{1,\dots,n\}$. We use $n$ for the number of variables in a constraint satisfaction problem, and $m$ for the number of constraints (edges in the case of colouring problems, clauses in the case of $k$-SAT). In the case of the graph colouring problem, we also write $r = \lceil \log (k+1) \rceil$, $s = \lceil \log (n+1) \rceil$. These represent the number of bits required to store an element of $[k] \cup \{\ast\}$, $\{0,\dots,n\}$ respectively.


\section{Grover's algorithm}
\label{sec:grover}

Given access to an oracle function $f:[N] \rightarrow \{0,1\}$, Grover's quantum search algorithm can be used to find $x$ such that $f(x)=1$, or output that no such $x$ exists, using $O(\sqrt{N})$ evaluations of $f$~\cite{grover97,boyer98}, with arbitrarily small failure probability $\delta$. The algorithm is based on ``Grover iterations'', each of which can be written as $DO_f$, where $D$ is a fixed ``diffusion operator'' and  $O_f$ is an oracle operator performing the map $\ket{x}\ket{y} \mapsto \ket{x}\ket{y \oplus f(x)}$. If the size $S$ of the set $\{x:f(x)=1\}$ is known in advance, the optimal number of Grover iterations to maximise the success probability can be calculated in advance; otherwise, one can show that running the algorithm using varying numbers of iterations (e.g.\ exponentially increasing, or random) is sufficient to solve the unstructured search problem. A precise analysis by Zalka~\cite{zalka99a} of one variant of the algorithm showed that, to achieve failure probability $\delta$, it is sufficient to carry out at most
\be \label{eq:grover} 1.582 \sqrt{N} \ln 1/\delta \ee
iterations\footnote{This is the ``simple algorithm'' in~\cite{zalka99a}; a different algorithm presented in~\cite{zalka99a} would be more efficient for small $\delta$.}. This is close to optimal, as even under the promise that $S=1$, $\Omega(\sqrt{N})$ evaluations of $f$ are required to find the unique $x$ such that $f(x)=1$ with high probability~\cite{bennett97,zalka99}. Here we will choose $\delta = 0.1$, where we obtain an upper bound of $3.642 \sqrt{N}$ iterations. (For this value of $\delta$, a lower bound of about $0.625\sqrt{N}$ iterations can be derived from the tight bound for the special case $S=1$, also shown by Zalka~\cite{zalka99}.)

Assuming that $N=2^n$ for some integer $n$ (as is the case for $k$-SAT), the diffusion operation can be implemented using a layer of Hadamard gates on every qubit, followed by a Toffoli gate controlled on all $n$ bits, with target an ancilla bit in the state $\frac{1}{\sqrt{2}}(\ket{0} - \ket{1})$, and then another layer of Hadamard gates. In most cases, the majority of the complexity in the algorithm therefore comes from the purely classical oracle operation $f$, as a Toffoli gate controlled on $n$ bits can be implemented using a circuit with $O(n)$ gates and depth $O(\log n)$.

\boxalgm{alg:grover}{Check whether $x$ violates any clause in a $k$-SAT formula $\phi$ with $m$ clauses.}{
{\bf Input:} $x \in \{0,1\}^n$.\\
{\bf Ancillae:} $m$-qubit register $A$.

\begin{enumerate}
\item Fan-out $x$ to $m$ copies.
\item In parallel, for each clause $c$: set $A_c$ to 1 if $x$ satisfies clause $c$.
\item Set the output bit to 1 if $A_c = 1$ for all $c$.
\item Uncompute $A$.
\item Fan-in $x$ back to 1 copy.
\end{enumerate}
}

In the case of $k$-SAT, the oracle needs to output 1 if and only if the input $x$ satisfies all clauses $c$ in $\phi$. This is an $m$-wise AND function of OR functions of $k$ bits each (and some additional NOT operations). There is a straightforward algorithm for implementing this depth-efficiently, which is stated as Algorithm \ref{alg:grover}. The key point is that to check all the clauses in parallel, $x$ needs to be fanned out to multiple copies, because two gates cannot be performed on the same qubit at the same time. Note that for random formulae, $m$ will usually be an overestimate of how many copies of each bit are required, because clauses that involve disjoint sets of variables can be checked at the same time. In the best case, the clauses could be grouped into approximately $mk/n$ sets of $n/k$ clauses, where the clauses in each set could be checked simultaneously. This would lead to $mk/n$ copies of each bit being required.

We can now calculate the complexity of Grover's algorithm for particular choices of $k$, $m$ and $n$. The algorithm consists of Toffoli gates and Clifford gates. We ultimately will be concerned with measuring the Toffoli depth and Toffoli count, with our primary goal being to minimise Toffoli depth (as this controls the runtime of the overall computation).

Each Toffoli gate controlled on $c \ge 2$ bits can be implemented using a circuit containing $2c-3$ standard Toffoli gates arranged into $2\lceil \log c \rceil - 1$ layers in a tree structure. However, almost half of these gates can be replaced with classically controlled Clifford gates using the ``uncompute AND'' operation described in~\cite{gidney17}, to give a circuit containing $c-1$ Toffoli gates. This does not change the measurement depth of the resulting circuit, as the classically controlled Clifford operation itself requires a measurement.


The operation of fanning out a single bit $b$ to $c$ copies can be implemented via a tree of CNOT operations of the following form, illustrated for $c=8$:
\[
\Qcircuit @C=0.3em @R=0.7em {
\lstick{\ket{b}} & \qw & \qw & \ctrl{1} &\qw&\qw&\qw&\qw& \ctrl{2} & \qw&\qw&\qw&\qw&\qw & \ctrl{4} & \qw & \qw & \qw & \qw & \qw & \qw& \rstick{\ket{b}} \\
\lstick{\ket{0}} & \qw & \qw & \targ&\qw&\qw& \qw & \qw & \qw & \ctrl{2}&\qw&\qw&\qw&\qw & \qw & \ctrl{4} & \qw & \qw & \qw & \qw & \qw& \rstick{\ket{b}} \\
\lstick{\ket{0}} & \qw & \qw & \qw&\qw&\qw& \qw & \qw & \targ&\qw&\qw&\qw&\qw & \qw & \qw & \qw & \ctrl{4}& \qw & \qw & \qw & \qw& \rstick{\ket{b}} \\
\lstick{\ket{0}} &\qw & \qw & \qw&\qw&\qw& \qw & \qw & \qw & \targ&\qw&\qw&\qw&\qw & \qw & \qw & \qw & \ctrl{4} & \qw & \qw & \qw& \rstick{\ket{b}}\\
\lstick{\ket{0}} & \qw & \qw &\qw&\qw&\qw& \qw & \qw & \qw&\qw&\qw&\qw&\qw & \qw & \targ & \qw & \qw & \qw &\qw & \qw & \qw& \rstick{\ket{b}}\\
\lstick{\ket{0}} & \qw & \qw &\qw&\qw&\qw& \qw & \qw & \qw&\qw&\qw&\qw&\qw & \qw & \qw & \targ & \qw & \qw &\qw & \qw & \qw& \rstick{\ket{b}}\\
\lstick{\ket{0}} & \qw & \qw &\qw&\qw&\qw& \qw & \qw & \qw&\qw&\qw&\qw&\qw & \qw & \qw & \qw & \targ & \qw &\qw & \qw & \qw& \rstick{\ket{b}} \\
\lstick{\ket{0}} & \qw & \qw &\qw&\qw&\qw& \qw & \qw & \qw&\qw&\qw&\qw&\qw & \qw & \qw & \qw & \qw & \targ & \qw & \qw & \qw & \rstick{\ket{b}}\\
}
\]
The depth of the circuit is $\lceil \log c \rceil$, and it uses $c-1$ CNOT gates and no other gates. However, in the surface code a fan-out operation (equivalently, a CNOT gate with multiple target bits) can be executed in the same time as required for one CNOT gate~\cite{fowler12}. So we assign this operation the same cost as one CNOT gate, which in any case is 0, when considering Toffoli depth and Toffoli count.

Combining all the components of Algorithm \ref{alg:grover}, the Toffoli depth is
\[ 4 \lceil (\log k) - 1 \rceil + 2 \lceil (\log m) - 1 \rceil + 3 \]
and the Toffoli count is $m(2k+1) - 3$.

The complexities of the various parts of the oracle operation are summarised in Table \ref{tab:complexities_grover} for a particular choice of parameters. It is obvious that the Toffoli depth is very low, while the Toffoli count seems very high in comparison. It is not clear that this could be reduced by more than an order of magnitude or so, though, given that there needs to be at least one Toffoli gate per clause (controlled on $k$ bits).

\begin{table}
\fontfamily{cmss}\selectfont
\resizebox{0.48\textwidth}{!}{
\begin{tabular}{|c|c|c|c|}
\hline Line & Operation & T-depth & Toffoli count \\
\hline 1, 5 & Fan-out / fan-in & 0 & 0 \\
\hline 2, 4 & Check clauses and uncompute & 14 & $2.3 \times 10^{7}$ \\
\hline 3 & AND of all clauses & 39 & $8.9 \times 10^{5}$ \\
\hline\hline & Total & 53 & $2.4 \times 10^7$\\
\hline
\end{tabular}
}
\caption{Representative complexities in the Grover oracle operation for $k=14$, $n=78$, $m=885743$.}
\label{tab:complexities_grover}
\end{table}


\section{Backtracking algorithms}
\label{sec:backtrack}

Some of the most effective and well-known classical algorithms for graph colouring and boolean satisfiability are based on the technique known as backtracking~\cite{davis62,davis60,sewell96,sansegundo12}. This approach can be applied to any constraint satisfaction problem where we have the ability to rule out partial solutions that have already failed to satisfy some constraints (e.g.\ partially coloured graphs where two adjacent vertices share the same colour). The basic concept behind this algorithm, in the context of graph colouring, is to build up a colouring by trying to assign colours to vertices. If a conflict is found, the colour is erased and a different colour is tried.

The skeleton of a generic backtracking algorithm that solves a $k$-ary CSP defined by a predicate $P$ and a heuristic $h$ is given in Algorithm \ref{alg:backtrack}. $P$ determines whether a partial assignment already violates a constraint, whereas $h$ determines the next variable to assign a value in a given partial assignment. Define $\mathcal{D} := ([k] \cup \{\ast\})^n$, where the $\ast$'s represent the positions which are as yet unassigned values.

\boxalgm{alg:backtrack}{General classical backtracking algorithm (see~\cite{montanaro18})}{
Assume that we are given access to a predicate $P: \mathcal{D} \rightarrow \{ \text{true}, \text{false}, \text{indeterminate} \}$, and a heuristic $h: \mathcal{D} \rightarrow \{1,\dots,n\}$ which returns the next index to branch on from a given partial assignment.
\\[3pt]
Return {\tt bt}$(\ast^n)$, where {\tt bt} is the following recursive procedure:
\\[3pt]
{\tt bt}$(x)$:
\begin{enumerate}
\item If $P(x)$ is true, output $x$ and return.
\item If $P(x)$ is false, or $x$ is a complete assignment, return.
\item Set $j = h(x)$.
\item For each $w \in [k]$:
\begin{enumerate}
\item Set $y$ to $x$ with the $j$'th entry replaced with $w$.
\item Call {\tt bt}$(y)$.
\end{enumerate}
\end{enumerate}
}

Backtracking algorithms can be seen as exploring a tree whose vertices correspond to partial solutions to the CSP. In the case where no solution is found, the runtime of the algorithm is determined by the size $T$ of this tree (along with the cost of computing $P$ and $h$). A quantum algorithm was described in~\cite{montanaro18} which solves the CSP with bounded failure probability in time scaling with $\sqrt{T}$, up to some lower-order terms. Two variants of the algorithm were given, with differing runtimes and preconditions. The first detects the existence of a solution and requires an upper bound on $T$. The second outputs a solution (if one exists) and does not require an upper bound on $T$; the price paid is a somewhat longer runtime.

\begin{thm}[\cite{montanaro18}]
\label{thm:detectbound}
Let $T$ be an upper bound on the number of nodes in the tree explored by Algorithm~\ref{alg:backtrack}. Then there is a quantum algorithm which, given $T$, evaluates $P$ and $h$ $O(\sqrt{Tn})$ times each, outputs true if there exists $x$ such that $P(x)$ is true, and outputs false otherwise. The algorithm uses $\poly(n)$ space, $O(1)$ auxiliary operations per use of $P$ and $h$, and fails with probability at most $0.01$.
\end{thm}

\begin{thm}[\cite{montanaro18}]
\label{thm:findbound}
Let $T$ be the number of nodes in the tree explored by Algorithm~\ref{alg:backtrack}. Then there is a quantum algorithm which makes $O(\sqrt{T}n^{3/2} \log n)$ evaluations of each of $P$ and $h$, and outputs $x$ such that $P(x)$ is true, or ``not found'' if no such $x$ exists. If we are promised that there exists a unique $x_0$ such that $P(x_0)$ is true, there is a quantum algorithm which outputs $x_0$ making $O(\sqrt{Tn} \log^3 n)$ evaluations of each of $P$ and $h$. In both cases the algorithm uses $\poly(n)$ space, $O(1)$ auxiliary operations per use of $P$ and $h$, and fails with probability at most $0.01$.
\end{thm}

These algorithms are based on the use of a quantum walk algorithm of Belovs~\cite{belovs13,belovs13a} to search efficiently within the backtracking tree. Their failure probabilities can be reduced to $\delta$, for arbitrarily small $\delta > 0$, at the cost of a runtime penalty of $O(\log 1/\delta)$.

There have been some subsequent improvements to these algorithms. First, in the case where the classical algorithm finds a solution without exploring the whole tree, its runtime could be substantially lower than $T$. Ambainis and Kokainis~\cite{ambainis17} showed that the quantum runtime bound can be improved to $\widetilde{O}(\sqrt{T'} n^{3/2})$, where $T'$ is the number of nodes actually explored by the classical algorithm. Second, Jarret and Wan~\cite{jarret18} showed that the runtime of Theorem \ref{thm:findbound} can be improved to $\widetilde{O}(\sqrt{Tn})$ without the need for any promise on the uniqueness of the solution\footnote{Their result is in fact stronger than this, as they show that $n$ can be replaced with a quantity depending on the maximum ``effective resistance'' of subtrees, which is upper-bounded by $n$ but might be smaller. On the other hand, the $\widetilde{O}$ notation hides a term logarithmic in the number of solutions, which can be very large.}. Also, the quantum backtracking algorithm has been applied to exact satisfability problems~\cite{mandra16}, the Travelling Salesman Problem~\cite{moylett17}, and attacking lattice-based cryptosystems~\cite{alkim16,delpino16}. However, none of these works precisely calculates the complexity of the algorithm for specific instances.

Filling in the details of the backtracking algorithmic skeleton requires implementing the $P$ and $h$ operations. We now describe how this can be done in the cases of $k$-colouring and satisfiability:

\begin{itemize}
\item $k$-colouring: $x \in \{([k] \cup \{\ast\})^n \}$ represents a partial $k$-colouring of $G$. $P(x)$ returns true if $x$ is a complete, valid colouring of $G$; false if there is a pair of adjacent vertices in $G$ that are assigned the same colour by $x$; and indeterminate otherwise. One natural choice for the heuristic $h$ is to choose the uncoloured vertex which is the most constrained (``saturated''), i.e.\ the one which has the largest number of adjacent vertices with different colours~\cite{brelaz79}.

\item $k$-SAT: $x \in \{0,1,\ast\}^n$ represents a partial assignment to variables in the formula. $P(x)$ returns true if $x$ is a complete, valid assignment; false if there exists a clause that $x$ does not satisfy; and indeterminate otherwise. A simple choice of $h$ is to order the variables in advance in some sensible way, and then to return the lowest index of a variable that has not yet been assigned a value. Here we ordered variables in decreasing order of the number of times they appear in the formula. 
\end{itemize}

In the case of $k$-SAT, one could also consider dynamic strategies for $h$ (e.g.\ choosing the variable that occurs in the largest number of clauses in the formula produced by substituting in the assigned values to variables, then simplifying). However, although these could give improved complexities for large instances, they seem likely to lead to larger runtime overheads per operation, so we did not consider them here. Conversely, in the case of $k$-colouring, one could consider static strategies for $h$ (e.g.\ choosing the highest-degree uncoloured vertex first). In our experiments, these seemed to be less efficient.

Undertaking a full analysis of the complexity of the backtracking algorithm then requires calculating the complexity of the $P$ and $h$ operations in detail, together with the complexity of the remaining operations in the algorithm. We do this in the next section.


\section{Backtracking algorithm complexity optimisation}
\label{sec:btopt}

For simplicity in the analysis, and to allow for future theoretical developments that may improve the runtime of the algorithm, we consider the simplest version of the backtracking algorithm of~\cite{montanaro18}: the algorithm that detects the existence of a marked vertex, given an upper bound on the number of vertices $T$ in the backtracking tree. (In practice, it may not be realistic to have access to such an upper bound; however, given a known distribution on problem instances, one could choose a bound that is expected to hold for most instances, for example.)

The algorithm is based on the use of a quantum walk to detect a marked vertex within a tree containing $T$ nodes. A marked node corresponds to a valid solution. Abstractly, the quantum walk operates on the Hilbert space $\mathcal{H}$ spanned by $\{\ket{r}\} \cup \{ \ket{x} :  x \in \{1,\dots,T-1\} \}$, where $r$ labels the root. The walk starts in the state $\ket{r}$. Let $A$ be the set of nodes an even distance from the root (including the root itself), and let $B$ be the set of nodes at an odd distance from the root. We write $x \rightarrow y$ to mean that $y$ is a child of $x$ in the tree. For each $x$, let $d_x$ be the degree of $x$ as a vertex in an undirected graph.

The walk is based on a set of diffusion operators $D_x$, where $D_x$ acts on the subspace $\mathcal{H}_x$ spanned by $\{\ket{x}\} \cup \{ \ket{y}: x \rightarrow y \}$. The diffusion operators are defined as follows:
\begin{itemize}
\item If $x$ is marked, then $D_x$ is the identity.
\item If $x$ is not marked, and $x \neq r$, then $D_x = I - 2 \proj{\psi_x}$, where
\[ \ket{\psi_x} = \frac{1}{\sqrt{d_x}} \left( \ket{x} + \sum_{y, x \rightarrow y} \ket{y} \right). \]
\item $D_r = I - 2 \proj{\psi_r}$, where
\[ \ket{\psi_r} = \frac{1}{\sqrt{1+ d_r n}} \left( \ket{r} + \sqrt{n} \sum_{y, r \rightarrow y} \ket{y}\right). \] 
\end{itemize}
A step of the walk consists of applying the operator $R_B R_A$, where $R_A = \bigoplus_{x \in A} D_x$ and $R_B = \proj{r} + \bigoplus_{x \in B} D_x$.

Then the detection algorithm from~\cite{montanaro18} is presented as Algorithm \ref{alg:detect}.

\boxalgm{alg:detect}{Detecting a marked vertex~\cite{montanaro18}}{
{\bf Input:} Operators $R_A$, $R_B$, a failure probability $\delta$, upper bounds on the depth $n$ and the number of vertices $T$. Let $\beta, K, L > 0$ be universal constants to be determined.
\begin{enumerate}
\item Repeat the following subroutine $K$ times:
\begin{enumerate}
\item Apply phase estimation to the operator $R_B R_A$ on $\ket{r}$ with precision $\beta/\sqrt{Tn}$.
\item If the eigenvalue is 1, accept; otherwise, reject.
\end{enumerate}
\item If the number of acceptances is at least $L$, return ``marked vertex exists''; otherwise, return ``no marked vertex''.
\end{enumerate}
}

We will mostly be interested in the complexity of the phase estimation step. The outer repetition step can be performed across multiple parallel runs of the algorithm, and hence does not affect the overall runtime (circuit depth). However, it does affect the overall cost of executing the algorithm, so we give an estimate of $K$ below.


\subsection{Optimisation of phase estimation step}

We first observe that the full phase estimation procedure is not actually required in Algorithm \ref{alg:detect}; it is sufficient to distinguish between the eigenvalue 1 and eigenvalues far from 1. This holds by the following claim:

\begin{claim}[See~\cite{montanaro18}, proof of Lemma 2.4; also see~\cite{belovs13,belovs13a}]
\label{claim:evs}
If there is a marked vertex, there exists an eigenvector $\ket{\phi}$ of $R_B R_A$ with eigenvalue 1 such that $|\braket{\phi|r}|^2 \ge 1/2$. Otherwise, $\| P_\chi \ket{r}\| \le \chi \sqrt{Tn}$ for any $\chi \ge 0$, where $P_\chi$ is the projector onto the span of eigenvectors of $R_B R_A$ with eigenvalues $e^{2 i \theta}$ such that $|\theta| \le \chi$.
\end{claim}

To distinguish between these two cases, we can perform the following algorithm, defined in terms of an integer $m$ to be determined:
\begin{enumerate}
\item Attach an ancilla register of $m$ qubits, initially in the state $\ket{0}^{\otimes m}$, to $\ket{r}$.
\item Apply $H^{\otimes m}$ to the ancilla qubits.
\item Apply the operation $\sum_{x = 0}^{M-1} \proj{x} \otimes (R_B R_A)^x$ using controlled-$R_B R_A$ gates, where $M=2^m$.
\item Apply $H^{\otimes m}$ to the ancilla qubits and measure them.
\item Accept if the outcome is $0^m$.
\end{enumerate}
To achieve sufficient precision, it is enough to take $M = O(\sqrt{Tn})$~\cite{cleve98a}; we will calculate a more precise expression for $M$ below. The quantum part of the algorithm is the same as the standard phase estimation algorithm~\cite{cleve98a}, but with the final inverse quantum Fourier transform replaced with Hadamard gates, which are negligible when calculating the overall complexity of the algorithm. The total number of controlled-$R_B R_A$ operations used is $M-1$. Note that step 3 of the algorithm can be implemented using singly-controlled operations, as in the standard phase estimation algorithm, because $M$ is a power of 2. A controlled-$(R_B R_A)^{2^{k-1}}$ gate controlled on the $k$'th qubit, for each $k$, will implement the desired transformation. Expanding $\ket{r} = \sum_j \alpha_j \ket{\phi_j}$ in terms of a basis $\ket{\phi_j}$ of eigenvectors of $R_B R_A$ with corresponding eigenvalues $e^{2 i \theta_j}$, where $\ket{\phi_0}$ corresponds to eigenvalue 1, the final state before the measurement is
\[ \frac{1}{\sqrt{M}} \sum_j \alpha_j \left( \sum_{x=0}^{M-1} e^{2 i \theta_j x} \ket{x} \right) \ket{\phi_j}, \]
so the probability $p$ that the algorithm accepts is
\[ \frac{1}{M^2} \sum_j |\alpha_j|^2 \left| \sum_{x=0}^{M-1} e^{2i\theta_j x} \right|^2 =: \sum_j |\alpha_j|^2 \mu_j. \]
We see that, if $\ket{r}$ were an eigenvector with eigenvalue 1 (i.e.\ $\alpha_0 = 1$) the algorithm would accept with certainty. If there is a marked element, $|\braket{\phi_0|r}|^2 \ge 1/2$ by Claim \ref{claim:evs}, so the algorithm accepts with probability at least $1/2$.

In the case where there is no marked element, we split up the sum over $k$ to obtain
\be \label{eq:p} p =  \sum_{j,|\theta_j| \le \chi} |\alpha_j|^2 \mu_j + \sum_{j,|\theta_j| > \chi} |\alpha_j|^2 \mu_j. \ee
Upper-bounding these sums now proceeds via a similar argument to standard calculations for phase estimation~\cite{cleve98a}, but we will attempt to find somewhat tighter bounds. We can evaluate
\[ \mu_j = \frac{|1-e^{2iM\theta_j}|^2}{M^2 |1-e^{2i\theta_j}|^2} = \frac{\sin^2(M\theta_j)}{M^2 \sin^2 \theta_j} \]
by the formula for a geometric series. Clearly $\sin^2(M\theta_j) \le 1$, and using $\sin x \ge x - x^3/6$ we have
\[ \mu_j \le \frac{1}{M^2 \theta_j^2 (1 - \theta_j^2/6)}. \]
We can now upper-bound $p$ as
\[ p \le \chi^2 Tn + \frac{1}{M^2 \chi^2(1-\chi^2/6)},  \]
where we upper-bound the first sum in (\ref{eq:p}) using Claim \ref{claim:evs} to infer that $\sum_{j,|\theta_j| \le \chi} |\alpha_j|^2 \le \chi^2 T n$, and that $\mu_j \le 1$; and upper-bound the second sum using $\sum_j |\alpha_j|^2 = 1$ and that the upper bound on $\mu_j$ decreases with $\theta_j$ in the range $[0,\pi/2]$.

We now optimise $\chi$ to find the best possible bound on $p$. Assume that $\chi = a / \sqrt{Tn}$, $M = \sqrt{Tn} / b$ for some constants $a$, $b$. Then the upper bound on $p$ becomes
\[ p \le a + \frac{b}{a(1 - a^2 / (6 \sqrt{Tn}))} = a + \frac{b}{a(1-o(1))}. \]
The minimum over $a$ of $a+b/a$ is $2\sqrt{b}$, so we obtain an overall bound that $p \le 2\sqrt{b}(1+o(1))$. In order to achieve a separation from $1/2$ in this case, it is sufficient to take $b < 1/16$; different choices of $b$ allow a tradeoff between the complexity of the phase estimation step, and the number of times it needs to be repeated in Algorithm \ref{alg:detect}. For the calculations here, we choose $b = 1/32$. Given this, we can now calculate (numerically) the value of $K$ in Algorithm \ref{alg:detect} required to obtain a desired probability of success. For $b =1/32$, to achieve failure probability $\delta = 0.1$ it is sufficient to take $K = 79$. In this case the overall cost of the algorithm, in terms of uses of controlled-$R_B R_A$, is at most
\be \label{eq:overallcost} \frac{K}{b} \sqrt{Tn} \le 2528 \sqrt{Tn}, \ee
and the number of sequential uses of $R_B R_A$ in the circuit is at most $32 \sqrt{Tn}$. The above algorithm assumes that $M$ is a power of 2, but the algorithm can easily be modified to handle the case where it is not, by replacing the use of $H^{\otimes m}$ with an operation creating a uniform superposition over $M < 2^m$ elements. The cost of this operation and its inverse is negligible in the context of the overall algorithm.

Note that the ``optimal'' variant of phase estimation where a different input state is used (rather than $\ket{+}^{\otimes m}$) does not seem to achieve a significantly better bound when $m$ is close to its minimal value (see e.g.\ the analysis in~\cite{childs10c}).


\subsection{Efficient parallel implementation of $R_A$ and $R_B$}

\boxalgm{alg:optbt}{$R_A$ operation for $k$-colouring (optimised and parallelised version of algorithm in~\cite{montanaro18}). Details of how the number of colours is stored are given in Section \ref{sec:misc}. Figure \ref{fig:ra} gives an illustrative circuit diagram.}{
{\bf Input:}\ Computational basis state $\ket{\ell}\ket{i_1,v_1}\ket{i_2,v_2}\dots\ket{i_\ell,v_\ell}\ket{0,\ast}\dots\ket{0,\ast}$, which corresponds to an input assigning value $v_j$ to the $i_j$'th variable.\\
{\bf Ancillae:} $x \in \mathcal{D}$, $a \in \{\ast\}\cup [k]$, $p \in \{$T, F, ?$\}$, $h \in \{0,\dots,n\}$, $c$.

\begin{enumerate}
\item Convert input to $x \in \mathcal{D}$, stored in ancilla register. If $\ell$ is odd, ignore the pair $(i_\ell,v_\ell)$ and instead swap $i_\ell$ with $h$ and swap $v_\ell$ with $a$.
\item Fan-out $x$ to $n(k+1)$ copies.
\item In parallel: Evaluate the number of colours used in $x$, stored in $c$ register, and if $\ell$ is even, evaluate $P(x)$, stored in $p$ register; if $\ell$ is even, evaluate $h(x)$, stored in $h$ register;  
\item If $a \neq \ast$, subtract 1 from $\ell$.
\item If $p=$ F, invert the phase of the input. 
\item If $p=$ ?, and $\ell > 0$, apply diffusion map $D$ to $a$ to mix over $c+1$ colours; if $p=$?, and $\ell = 0$, apply diffusion map $D'$ to $a$ to mix over $c+1$ colours.
\item If $a \neq \ast$, add 1 to $\ell$.
\item In parallel: Uncompute $c$ and $p$; if $a = \ast$, uncompute $h$.
\item Fan-in $x$ back to 1 copy.
\item Unconvert $x$. If $\ell$ is odd, ignore the pair $(i_\ell,v_\ell)$ and instead swap $a$ with $v_\ell$ and swap $h$ with $i_\ell$.
\end{enumerate}
}

To implement the $R_A$ and $R_B$ operations efficiently requires some additional work. A full description of how these operations can be implemented was given in~\cite{montanaro18} (Algorithm 3). Here, in Algorithm \ref{alg:optbt} we describe an essentially equivalent algorithm for the case of $k$-colouring which can be implemented more efficiently, and which computes operations in parallel where possible. We illustrate the algorithm in Figure \ref{fig:ra} at the end of the paper. The algorithm as used for $k$-SAT is similar, but simpler, and the modifications required for this are described in Section \ref{sec:btksat}.

In Algorithm \ref{alg:optbt}, $D$ corresponds to inversion about the state $\ket{\psi} = \sum_{i \in \{\ast\} \cup [c+1]} \ket{i}$, and $D'$ corresponds to inversion about a state $\ket{\psi'}$ of the form $\ket{\psi'} = \alpha \ket{\ast} + \frac{\beta}{\sqrt{k}} \sum_{i \in [c+1]} \ket{i}$. The algorithm implements $R_A$; the operation $R_B$ is similar, but slightly simpler because the alternative diffusion map $D'$ does not need to be performed. Correctness can be checked by tracking the effect of the algorithm on a computational basis state; we omit the routine details.

The main points of comparison between Algorithm \ref{alg:optbt} and the algorithm presented in~\cite{montanaro18} are:

\begin{itemize}
\item Both algorithms need to convert between two input representations: one of the form $(i_1,v_1),\dots,(i_\ell,v_\ell)$, representing that variable $i_j$ is assigned value $v_j$, and one of the form $x \in \mathcal{D}$, representing a partial assignment with $\ast$'s indicating unassigned variables. This is necessary because the first form allows efficient assignment of a value to a variable, while the second allows efficient checking against constraints. In Algorithm \ref{alg:optbt}, this conversion is performed once for both $P$ and $h$, as opposed to being done internally to each of $P$ and $h$ separately.

\item The most complicated operations in the algorithm are $P$ and $h$. In Algorithm \ref{alg:optbt}, they are executed in parallel, and each of them in turn contains some operations performed in parallel. To achieve this, copies of the input need to be produced using fan-out operations.

\item At each step of the algorithm, it is necessary to diffuse over neighbours of nodes in the backtracking tree. The algorithm in~\cite{montanaro18} achieves this by checking whether $P(x')$ is false for each of the children $x'$ of the current node $x$ in turn, and only diffusing over those children for which this is not the case. For simplicity and efficiency, Algorithm \ref{alg:optbt} defers this check to when the child nodes themselves are explored.

\item The standard backtracking algorithm would have $k$ children for each node in the backtracking tree, corresponding to the $k$ colours available. However, to determine colourability, it is more efficient (and equivalent) to only allow the algorithm to choose a colour between 1 and $c+1$ for each subsequent node to be coloured, where $c$ is the number of colours currently used. In particular, this enforces the constraint that all colourings considered include all colours between 1 and $c'$, for some $c'$. This optimisation is used in the classical DSATUR algorithm.
\end{itemize}

It remains to describe the steps of Algorithm \ref{alg:optbt} in more detail, including the implementation of the $P$ and $h$ operations for particular problems. The overall complexities of the various steps found are summarised in Table \ref{tab:complexities_bt_kcol} and \ref{tab:complexities_bt_ksat} in the cases of graph colouring and SAT, respectively, for choices of parameters close to the limit of problem size that can be solved in one day. The total scaling of the T-depth of $R_A$ with problem size in the case of graph colouring is illustrated in Figure \ref{fig:overhead}.

\begin{table}
\fontfamily{cmss}\selectfont
\resizebox{0.48\textwidth}{!}{
\begin{tabular}{|c|c|c|c|}
\hline Line & Operation & T-depth & T/Toffoli count \\
\hline 1, 10 & Conversion & 160 & $1.7 \times 10^{7}$\\
\hline 2, 9 & Fan-out / fan-in & 2 & $1.4 \times 10^{3}$\\
\hline 3 & Compute $c$ & 169 & $1.5 \times 10^{3}$ \\
& Compute $P(x)$ & 88 & $5.4 \times 10^{5}$\\
& Compute $h(x)$ & 903 & $3.6 \times 10^{6}$\\
\hline 6 & Diffusion & 1032 & $3.9 \times 10^{3}$\\
\hline 8 & Uncompute $c$ & 169 & $1.5 \times 10^{3}$ \\
& Uncompute $P(x)$ & 80 & $3.0 \times 10^{5}$\\
& Uncompute $h(x)$ & 903 & $3.6 \times 10^{6}$\\  
\hline 4, 5, 7 & Other operations & 114 & $3 \times 10^2$ \\
\hline\hline & Total & 3114 & $2.5 \times 10^{7}$\\
\hline
\end{tabular}
}
\caption{Representative complexities in the $R_A$ operation for $k$-colouring with $n=136$, $k=19$.}
\label{tab:complexities_bt_kcol}
\end{table}

\begin{table}
\fontfamily{cmss}\selectfont
\resizebox{0.48\textwidth}{!}{
\begin{tabular}{|c|c|c|c|}
\hline Line & Operation & T-depth & T/Toffoli count \\
\hline 1, 10 & Conversion & 144 & $2.5 \times 10^{6}$ \\
\hline 2, 9 & Fan-out / fan-in & 2 & $2.8 \times 10^{2}$ \\
\hline 3 & Compute $P(x)$ & 87 & $1.0 \times 10^{7} $ \\
& Compute $h(x)$ & 57 & $2.9 \times 10^{2}$ \\
\hline 6 & Diffusion & 98 & $1.4 \times 10^{2}$ \\
\hline 8 & Uncompute $P(x)$ & 87 & $9.7 \times 10^{6}$ \\
& Uncompute $h(x)$ & 57 & $2.9 \times 10^{2}$ \\  
\hline 4, 5, 7 & Other operations & 106 & $2.9 \times 10^2$ \\
\hline\hline & Total & 524 & $2.3 \times 10^{7}$ \\
\hline
\end{tabular}
}
\caption{Representative complexities in the $R_A$ operation for $k$-SAT with $k=12$, $n=71$, $m=201518$.}
\label{tab:complexities_bt_ksat}
\end{table}


\subsection{Calculating circuit complexities}
\label{sec:basic}

With the exception of step 6, all of the operations in $R_A$ are completely classical, and can be described in terms of Toffoli gates controlled on $m \ge 0$ bits (incorporating NOT and CNOT gates). We use the same depth-efficient construction of Toffoli gates controlled on $m$ bits that was discussed in the context of Grover's algorithm. When $R_A$ is used, it is a controlled operation itself, so we must add 1 to the number of control lines used by all of its gates.
In particular, to compute the T-depth of the overall circuit we need to keep track of the ``CNOT depth'' of $R_A$. In the description of $R_A$ below, for readability we do not include the additional control line required, but we do include it when discussing T-depth and Toffoli count. Also note that in some cases below, we do not give formulae for the T-depth and Toffoli count, but compute this via a program, available at~\cite{code}.

Throughout the implementation, we represent an element of the set $[k] \cup \{ \ast \}$ as a $r := \lceil \log(k+1) \rceil$-bit string, where $\ast$ corresponds to $0^r$. The ?, F, T values that the $p$ register can take are represented as 00, 01, 10, respectively. Where algorithms for steps state usage of ancillas, this only encompasses those required to describe the algorithm (i.e.\ subroutines within the steps may use additional ancillas).


\subsection{Conversion of input}

In the first step of the algorithm, we need to convert an input of the form $(i_1,v_1), \dots, (i_n,v_n)$ to a string $x \in \mathcal{D}$. We also need to ignore a particular pair $(i_\ell,v_\ell)$ if $\ell$ is odd. An algorithm to do all of this is described as Algorithm \ref{alg:convert}. The main idea behind the algorithm is to create an $n \times n$ array $B$ where $B_{pq}$ stores $v_p$ if $i_p = q$, and is otherwise zero (so, for each $p$, $B_{pq}$ is nonzero for at most one $q$). Then, for each $q$, if such a nonzero element $B_{pq}$ exists it is copied to $x_q$.

\boxalgm{alg:convert}{Conversion of input.}{
{\bf Input:} $n$ pairs $(i_j, v_j)$, $h$, $a$; $i_j,h \in \{0,\dots,n\}$, $v_j,a \in [k] \cup \{\ast\}$.\\
{\bf Output:} $x \in \mathcal{D}$.\\
{\bf Ancillae:} $n \times (n-1) \times (r+s)$-qubit register $A$; $n \times n \times r$-qubit register $B$.
\begin{enumerate}
\item Fan-out each of the $n$ pairs $i_j,v_j$ to $n$ copies, stored in register $A$ and the input register.
\item In parallel, for each pair $p,q \in [n]$:
\begin{enumerate}
\item If $\ell$ is even or $\ell \neq p$, and $i_p = q$, copy $v_p$ to $B_{pq}$;
\item If $\ell$ is odd and $\ell = p$, swap $i_p$ with $h$ and swap $v_p$ with $a$.
\end{enumerate}
\item In parallel, for each $q$: copy $\sum_p B_{pq}$ to $x_q$.
\item Uncompute $B$ and then $A$.
\end{enumerate}
}

We now describe each of the operations in Algorithm \ref{alg:convert} in more detail, and calculate their complexities. First, the fan-out and fan-in operations are implemented in the same way as in Grover's algorithm. In step 1 of Algorithm \ref{alg:convert}, each bit can be fanned out in parallel, so this operation corresponds to $n(r+s)$ parallel fan-outs of 1 bit to $n$ bits.

To implement step 2, we can first use a Toffoli gate controlled on $s$ bits with a target of one ancilla bit to store whether $\ell = p$ (in the case of odd $p$; for even $p$, this is omitted). Then to carry out step 2a, we perform $r$ parallel bit copy operations controlled on this bit being 0 and on $i_p = q$ (corresponding to a Toffoli controlled on $s+2$ bits). Step 2b then consists of two controlled-swap operations, each controlled on the ancilla bit, which can be performed in parallel. Swapping pairs of bits can be implemented as 3 CNOT gates. At the end of step 2, we uncompute the ancilla bit.

For step 3, we need to copy the one non-zero element of the $n$-element set $S_q := \{B_{pq'} : q' = q\}$ to $x_q$, for each $q$. We could achieve this by a binary tree of addition operations (qv) but a simpler method is to use a tree of CNOT gates. We consider each set $S_q$ in parallel, and each bit within elements of this set in parallel too, corresponding to a sequence of $n$ bits containing at most one 1. For sequences of this form, their sum is just the same as their sum mod 2. So, for each pair of bits $(b_i,b_j)$, we can copy their sum into an ancilla qubit with the circuit
\[
\Qcircuit @C=0.5em @R=1em {
\lstick{\ket{b_i}} & \ctrl{2} & \qw & \qw & \rstick{\ket{b_i}} \\
\lstick{\ket{b_j}} & \qw & \ctrl{1} & \qw  & \rstick{\ket{b_j}}  \\
\lstick{\ket{0}} & \targ & \targ & \qw  & \rstick{\ket{b_i \oplus b_j}}\\
}
\]
Using a binary tree of such addition operations, we can set each bit of $x_q$ correctly in depth $4 \lceil \log n \rceil$ (including the cost of uncomputing to reset the ancillas), where all gates are CNOT gates. As precisely $n-1$ bitwise summation operations take place, the circuit uses $n-1$ ancillas and $4(n-1)$ CNOT gates for each bit in each $q$. These quantities are multiplied by $n r$ to obtain the overall ancilla and gate complexities.

In step 4 we finally need to uncompute the $B$ and $A$ registers, with the same complexities as computing them.

For large $n$, the T-depth of the circuit is dominated by steps 1, 3 and 4, and is approximately $6 \lceil \log n \rceil$. This will turn out to be negligible in the context of the overall algorithm.


\subsection{Evaluation of $P(x)$: checking violation of a constraint}

The algorithm for evaluating the predicate $P$ in the case of graph colouring is presented as Algorithm \ref{alg:p}. The goal is to check whether, given a partial assignment of colours to vertices in a graph $G$, there exists an edge of $G$ whose endpoints are assigned the same colour. The output should be false if so, true if there is no such edge and all vertices are assigned a colour, and undetermined otherwise. We can hard-code the edges of $G$ into a quantum circuit for checking this, making the algorithm quite straightforward. Steps 1 and 2 check whether any constraint is violated, while steps 3 and 4 check whether the partial assignment is complete. As we have access to $n$ copies of $x$, we can perform all the checks in parallel for all edges, using a decomposition of the $m \le \binom{n}{2}$ edges into at most $n$ matchings, where the edges in each matching can be checked in parallel.

\boxalgm{alg:p}{$P(x)$: Checking violation of a constraint.}{
{\bf Input:} $n$ copies of $x \in \mathcal{D}$.\\
{\bf Output:} ?, F, T (represented as 00, 01, 10)\\
{\bf Ancillae:} $m$-qubit register $A$, $n$-qubit register $B$.

\begin{enumerate}
\item In parallel: for each edge $e = (i,j)$ in $G$, check whether $x_i = x_j$ and $x_i \neq \ast$. Set $A_e$ to 1 if so (corresponding to edge $e$ being violated).
\item Set the second bit of the output register to 1 if any of the bits of $A$ are equal to 1.
\item In parallel: for each $i \in [n]$, set $B_i$ to 1 if $x_i = \ast$.
\item If the second bit of the output register is 0, and $B_i = 0$ for all $i$, set the first bit of the output register to 1.
\item Uncompute $B$ and $A$.
\end{enumerate}
}


In step 1, checking equality of 2 bits $a$, $b$ can be done using one ancilla and 2 CNOT gates, via the following circuit:
\[
\Qcircuit @C=0.5em @R=1em {
\lstick{\ket{a}} & \ctrl{2} & \qw & \qw & \rstick{\ket{a}} \\
\lstick{\ket{b}} & \qw & \ctrl{1} & \qw & \rstick{\ket{b}}  \\
\lstick{\ket{1}} & \targ & \targ & \qw & \rstick{\ket{a=b}} \\
}
\]
These equality checks can be done in parallel across all the bits of $x_i$, $x_j$, followed by an Toffoli gate controlled on the $r$ equality test bits.
The check $x_i \neq \ast$ corresponds to at least one of the bits of $x_i$ not being equal to 0, which can be checked using a similar Toffoli, some NOT gates, and one more ancilla. We then use one more Toffoli gate to write the AND of these into bit $A_e$. We do not need to uncompute the ancilla bits used in these steps, as we will do this in step 5.

Step 2 can be implemented using a Toffoli gate controlled on $m$ qubits; step 3 can be implemented via $n$ Toffoli gates in parallel, each controlled on $r$ bits; step 4 can be implemented using a Toffoli gate controlled on $n + 1$ bits (together with some NOT gates in each case).


\subsection{Evaluation of $h(x)$: choosing the next vertex}
\label{sec:h}

The algorithm for computing $h(x)$ in the case of $k$-colouring is presented as Algorithm \ref{alg:h}. The goal of the algorithm is to output the uncoloured vertex that is the most constrained or ``saturated'' (has the largest number of adjacent vertices assigned distinct colours).

\boxalgm{alg:h}{$h(x)$: Choosing the next vertex.}{
{\bf Input:} $kn$ copies of $x \in \mathcal{D}$.\\
{\bf Output:} Identity of the vertex labelled $\ast$ whose adjacent vertices are labelled with the largest number of different colours.\\
{\bf Ancillae:} $n \times n \times k$-qubit register $A$; $n$ $r$-qubit registers $B_i$.

\begin{enumerate}
\item In parallel: for each triple $(i,j,c)$ such that $i$ is adjacent to $j$, set $A_{ijc} = 1$ if $x_j = c$.
\item For each $i$, set $B_i = \sum_c \bigvee_j A_{ijc}$.
\item Copy $\max_i B_i$ into the output register.
\item Uncompute steps 2 and 1.
\end{enumerate}
}

Each of the operations performed in step 1 of the algorithm corresponds to a Toffoli gate controlled on $r$ bits (together with some NOT gates). Similarly to the case of the $P$ operation, they can all be performed in parallel, using copies of the input $x$.

The most complicated part is the second and third steps. The second step can be split into two parts. First, we initialise an ancilla register $C_{ic} = \bigvee_j A_{ijc}$; then we set $B_i = \sum_c C_{ic}$. Each bit $C_{ic}$ can be set in parallel using a Toffoli gate controlled on $n$ qubits (and some NOT gates). Summing these bits efficiently in parallel can be achieved via a binary tree of addition operations, where the tree is of depth $\lceil \log n \rceil$. The $t$'th level of the tree sums approximately $n/2^t$ pairs of integers in parallel, producing $(t+1)$-bit integers as output. The addition operations themselves can be carried out using remarkably efficient out-of-place addition circuits presented by Draper et al.~\cite{draper06}. Even for adding two 10-bit numbers, the depth of the circuit given in~\cite{draper06} is only 11 (8 layers of which are made up of Toffoli gates, the remainder of which contain CNOTs). Here this can be reduced further by simplifying the circuit in the earlier layers of the tree that compute smaller numbers, and by observing that some layers of the circuit in~\cite{draper06} are only used to uncompute ancilla bits, which are uncomputed anyway in step 4. Hence these layers can be omitted.


The third step can be achieved via a binary tree of ``compare-and-swap'' operations. The $n$ $B_i$ values are split into adjacent pairs. Then the efficient in-place comparator also presented by Draper et al.~\cite{draper06} is used to determine which element of each pair is larger. If they are out of order, they are swapped, using a decomposition of a controlled-SWAP operation in terms of 3 Toffoli gates. The result at the end of the tree is that the largest value is moved to the bottom, and we can then copy it into the output register. We then need to reverse the rest of step 3 to put the elements of $B_i$ back into their original order before uncomputing steps 1 and 2.


\subsection{Diffusion operations}
\label{sec:diffusion}

\boxalgm{alg:diffuse}{Diffusion in backtracking tree (assuming $k+1$ not a power of 2). $U_\ast$, $V_c$, $\alpha$, $\beta$ are defined in the text.}{
{\bf Input:} $a \in [k] \cup \{\ast\}$, $p \in \{\text{?, F, T}\}$, $\ell \in \{0,\dots,n\}$, $c \in \{0,\dots,k\}$.\\
{\bf Output:} Diffusion map applied to $a$.\\
{\bf Ancillae:} qubits $A$, $B$, $C$ initially in state $\ket{0}$.

\begin{enumerate}
\item Set $B$ to 1 if $\ell = 0$.
\item Set $A$ to 1 if $p = $ ? and $B = 0$.
\item If $A=1$, perform $D = V_c U_\ast V_c^\dag$ on $a$.
\item If $B=1$, set $C$ to $\alpha \ket{0} + \beta \ket{1}$.
\item If $B=C=1$, perform $D$ on $a$.
\item If $B=1$, perform a Hadamard gate on $C$.
\item If $B=C=1$ and $a \neq 0^r$, invert the phase of $a$.
\item If $B=1$, perform a Hadamard gate on $C$.
\item Uncompute $B$ and $A$.
\end{enumerate}
}

In step 6 of the $R_A$ operation, we need to apply a diffusion map $D$ or $D'$ to the state of an ancilla register $\ket{a}$ of $r$ qubits, controlled on some other qubits. See~\cite{cade18} for a general discussion of how this can be achieved in quantum walk algorithms. Recall that $D$ corresponds to inversion about the state $\ket{\psi} = \frac{1}{\sqrt{c+2}} \sum_{i \in \{\ast\} \cup [c+1]} \ket{i}$, and $D'$ corresponds to inversion about a state $\ket{\psi'}$ of the form $\ket{\psi'} = \alpha \ket{\ast} + \frac{\beta}{\sqrt{c+1}} \sum_{i \in [c+1]} \ket{i}$. Further recall that we represent $\ast$ as $0^r$. First assume that $c$ is fixed throughout (we will handle $c$ being given as input to the algorithm later).

The algorithm for achieving this map is described as Algorithm \ref{alg:diffuse}. We begin by using two ancilla qubits $A$ and $B$ to store whether (a) $p=$ ?, and $\ell > 0$; (b) $\ell = 0$. If we set $B$ first, these can be implemented using Toffoli and NOT gates controlled on 3 and $s$ qubits, respectively.
Controlled on $A$, we apply $D$; controlled on $B$, we apply $D'$. Afterwards, we uncompute the ancilla qubits. Note that, once the ancilla qubits have been set, we do not need to access the overall $R_A$ control bit again.

We first describe how to implement $D$. To do this, it is sufficient to implement $V$ such that $V\ket{\ast} = \ket{\psi}$. This is because the combined operation $V U_{\ast} V^\dag$, where $U_{\ast} \ket{\ast} = -\ket{\ast}$ and $U_{\ast} \ket{x} = \ket{x}$ if $x\neq\ast$, maps $\ket{\psi} \mapsto -\ket{\psi}$, $\ket{\psi^\perp} \mapsto -\ket{\psi^\perp}$ as desired. $U_\ast$ can be implemented using a Toffoli gate controlled on $r$ qubits and an ancilla qubit in the state $\ket{-} = \frac{1}{\sqrt{2}}(\ket{0}-\ket{1})$.

Implementing $V$ is easy if $c=k-1$ and $k+1$ is a power of 2 (and hence $\ket{\psi} = \ket{+}^{\otimes r}$), by setting $V = H^{\otimes r}$. In the context of the overall circuit, this corresponds to implementing $r$ copies of a controlled-Hadamard gate. An efficient Clifford+T circuit for this gate is given in~\cite{bera10} which has T-depth 2 and T-count 2, and uses no ancillas. The total resources required to implement $D$ are then only T-depth $2(\lceil \log(r-1) \rceil + 3$, T-count $10r - 9$, and $2(r-2)$ ancillas (excluding the ancilla in the state $\ket{-}$, which can be reused from step 5 of $R_A$).

If $c < k-1$ or $k+1$ is not a power of 2, we need to reflect about a state that is not $\ket{+}^{\otimes r}$. We can do this using just one iteration of the exact amplitude amplification trick of Brassard et al.~\cite{brassard02}. Let $i$ be the smallest integer such that $c + 2 \le 2^i$. Then define $H' = H'' \otimes I^{\otimes r-i} \otimes H^{\otimes i}$, where $H'' \ket{0} = \gamma \ket{0} + \delta \ket{1}$, and $|\gamma|^2 = 2^{i-2} / (c+2)$. Note that this is well-defined, because $c+2 \ge 2^{i-1}$. Then $V$ can be implemented using the following sequence of operations:
\[ - H' U_\ast (H')^\dag U_{\le c+1} H', \]
where $U_{\le c+1}$ is an operation which inverts the phase of any computational basis state corresponding to a value less than or equal to $c+1$, and which leaves all other computational basis states unchanged. By the standard analysis of amplitude amplification (where $U_{\le c+1}$ is the ``verifier'' and $H'$ is the guessing algorithm) this will produce the state $\ket{0}\ket{\psi}$ when applied to the state $\ket{0}\ket{0}$, as can be confirmed by direct calculation.
 
 Many of the operations in $V$ are quite efficient. To implement the less-than checking step $U_{\le c+1}$, we can use the efficient comparison circuit of Draper et al.~\cite{draper06}.
The other operations are (controlled) single-qubit gates. The only one which has not yet been discussed is $H''$. This is not straightfoward to implement, as it requires approximate synthesis using Clifford and T operations. An arbitrary operation mapping $\ket{0}$ to $\cos \theta \ket{0} + \sin \theta \ket{1}$ can be written as $THRH$, for some rotation $R$ in the Z plane. It was shown in~\cite{bocharov15} that the mean expected T-count to implement Z-rotations up to inaccuracy $\epsilon$ using ``repeat-until-success'' circuits is $\approx 1.15 \log_2(1/\epsilon) + 9.2$. In order to achieve a good overall level of accuracy for the whole algorithm, we require $\epsilon$ to be upper-bounded by the total number of uses of $D$ in the circuit. This is at most $64 \sqrt{Tn}$ by (\ref{eq:overallcost}) and the text below it. In the algorithm we actually need a controlled-$H''$ operation (controlled on $A$). Using a technique of Selinger~\cite{selinger13}, controlled-$R$ and controlled-$T$ can each be implemented at an additional cost of T-depth 2 and T-count 8. So the expected overall cost of implementing $H''$ is T-depth $\approx 1.15 \log_2(64 \sqrt{Tn}) + 17.2$, T-count $\approx 1.15 \log_2(64 \sqrt{Tn}) + 29.2$.

It remains to implement $D'$, which can be done as follows. First, add another ancilla qubit $C$ in the state $\ket{0}$ and map it to the state $\alpha \ket{0} + \beta \ket{1}$ using a similar synthesis technique. Then use a controlled-$D$ operation (controlled on $C$) to produce $\alpha \ket{0}\ket{\ast} + \beta \ket{1}\frac{1}{\sqrt{k}} \sum_{i \in [k]} \ket{i}$, followed by a Hadamard gate on $C$ to produce
\[ \frac{1}{\sqrt{2}} \big( \ket{0}\big(\alpha \ket{\ast} + \frac{\beta}{\sqrt{k}} \sum_{i \in [k]} \ket{i} \big) + \ket{1}\big(\alpha \ket{\ast} - \frac{\beta}{\sqrt{k}} \sum_{i \in [k]} \ket{i} \big) \big). \]
Next, conditional on $C$ being in the state 1 and the other qubits not being in the state $0^r$, the phase is inverted to produce the state
\[ \frac{1}{\sqrt{2}} \big( \ket{0} + \ket{1} \big) \big(\alpha \ket{\ast} + \frac{\beta}{\sqrt{k}} \sum_{i \in [k]} \ket{i} \big). \]
A final Hadamard gate on $C$ restores it to its initial state.

We finally describe how to handle dependence on $c$. When used in Algorithm \ref{alg:optbt}, we can think of steps 3--8 of the overall diffusion operation as being provided with an input of the form $\ket{c}\ket{A}\ket{B}\ket{a}$, where $c$ specifies the number of colours used in $x$. The goal is to apply an operation depending on $c$ (call this $D_c$) to $\ket{a'} := \ket{A}\ket{B}\ket{a}$. $D_c$ encompasses the controlled-$D$ and controlled-$D'$ operations above. This can be achieved by attaching $k-1$ ancilla registers of $r+2$ qubits each, where the $i$'th register is initially in a state $\ket{e_i}$ which is an eigenvector of $D_i$ with eigenvalue 1. These states can be prepared in advance of the algorithm, so the cost of preparing them is negligible in terms of the overall complexity. Then, before step 3 in Algorithm \ref{alg:diffuse}, swap operations controlled on $c$ are used to swap $\ket{a'}$ with $\ket{e_c}$, before applying $D_i$ to each register $i$ in parallel, and then swapping $\ket{a'}$ back into place. This has the effect of applying $D_c$ to $\ket{a'}$ and leaving the other registers unchanged.

Rather than applying $k-1$ controlled-swap operations sequentially, we can achieve a reduction in depth by storing all binary prefixes of the $r$-bit string corresponding to $c$. We can then use an $r$-step procedure where, at step $i$, the state $\ket{a'}$ is swapped into a register labelled with the first $i$ bits of $c$ (with the final register corresponding to $\ket{e_c}$). These swap operations can all occur in parallel for different choices of prefixes of length $i-1$. We also only need to perform swaps for the cases where $c_i = 1$, because at the $(i+1)$'th step we can associate each register labelled with an $i$-bit string with the corresponding $(i+1)$-bit string ending in 0. The final result is a ``controlled-swap tree'' of depth $r$.


\subsection{Miscellaneous parts and overall bound}
\label{sec:misc}

\boxalgm{alg:c}{Counting the number of colours used in $x$.}{
{\bf Input:} $k$ copies of $x \in \mathcal{D}$.\\
{\bf Output:} The number $c$ of colours used in $x$, and a bit-string $e$ checking equality to each prefix of the binary string corresponding to $c$.\\
{\bf Ancillae:} $n \times k$-qubit register $A$.

\begin{enumerate}
\item In parallel: for each $i$, $j$, set $A_{ij} = 1$ if $x_i = j$.
\item Set $c = \sum_j A_{ij}$.
\item For each possible prefix $p$ of $\{0,1\}^r$, check equality of the corresponding bits of $c$ to $p$, and store the answer in $e_p$.
\item Uncompute step 1.
\end{enumerate}
}

In step 3, we need to count the number $c$ of colours used in $x$, and also store whether each prefix of the binary string $c$ (i.e.\ each substring of the form $c_1\dots c_i$) is equal to each possible prefix (i.e.\ element of $\{0,1\}^i$). The algorithm for this is given as Algorithm \ref{alg:c}. The part that counts the number of colours used in $x$ is essentially a simplified version of Algorithm \ref{alg:h}. Then, to calculate the complexity of step 3 of Algorithm \ref{alg:c}, observe that there are $\sum_{i=1}^r 2^i = 2^{r+1}-1$ non-trivial prefixes of $x$, and checking each $i$-bit prefix can be achieved using a Toffoli gate controlled on $i$ bits.

In steps 2 and 9, we perform fan-out and fan-in operations, whose complexities can be calculated similarly to step 1. In steps 4 and 7, we need to perform a controlled decrement and increment, controlled on $a \neq \ast$. These can be done by first setting an ancilla qubit based on the $r$ bits of $a$, then using a controlled version of the in-place addition circuit of Draper et al.~\cite{draper06} (hardcoding the first input to 1 or $-1$), and then uncomputing the ancilla. We can reduce the complexity of the circuit stated in~\cite{draper06} slightly by restricting to 8-bit integers and observing that some of the gates can be removed because of the input hardcoding. The total depth is then 14 layers of Toffoli gates and 3 layers of CNOT gates, and the circuit uses 38 Toffoli gates and 25 CNOT gates.

The total T-depth of each of these two steps is thus at most
\[ 2 \times (2\lceil \log r \rceil + 1) + 3 \times 14 + 3 = 4 \lceil \log r \rceil + 47. \]
Finally, in step 5, we need to invert the phase of the input if $p=$ F. This can be done using a Toffoli gate and one ancilla qubit in the state $\frac{1}{\sqrt{2}}(\ket{0}-\ket{1})$. 

%
%


\section{Backtracking for $k$-SAT}
\label{sec:btksat}

The quantum backtracking algorithm can be applied to boolean satisfiability (SAT) problems in much the same way as for colouring problems. In the case of SAT, variables are boolean; within the backtracking algorithm, this corresponds to 2 bits being used to store each variable (we use $\ast \mapsto 00$, $F \mapsto 01$, $T \mapsto 10$). As the $P$ and $h$ operations are substantially simpler than in the case of graph colouring, the runtime of the algorithm is dominated by the diffusion operation, whose cost in turn is dominated by the gate synthesis step required. We can reduce the cost of this by observing that diffusion becomes much simpler if each variable represents an element of a set of size $2^c-1$, for some integer $c$, so allowing the two bits representing $x_i$ to take the fictitious value 11 makes this step more efficient. We must then modify the $P$ operation to check and reject this value. This increases the cost of the $P$ operation somewhat, and also increases the size of the backtracking tree by a factor of $3/2$. However, it is still advantageous overall. Note that some gate synthesis is still required to implement the $D'$ operation used in $R_A$.


\subsection{Evaluation of $P(x)$: checking violation of a constraint}
\label{sec:psat}

\boxalgm{alg:psat}{$P(x)$ for SAT: Checking violation of a constraint.}{
{\bf Input:} $m$ copies of $x \in \mathcal{D}$, where $x_i$ is represented by two bits.\\
{\bf Output:} ?, F, T (represented as 00, 01, 10)\\
{\bf Ancillae:} $(m+1)$-qubit register $A$, $n$-qubit register $B$.

\begin{enumerate}
\item In parallel: for each clause $c$ in $\phi$, check whether $x$ violates $c$. Set $A_c$ to 1 if so.
\item If $x_i = 11$ for any $i$, set $A_{m+1}$ to 1.
\item Set the second bit of the output register to 1 if any of the bits of $A$ are equal to 1.
\item In parallel: for each $i \in [n]$, set $B_i$ to 1 if $x_i = \ast$.
\item If the second bit of the output register is 0, and $B_i = 0$ for all $i$, set the first bit of the output register to 1.
\item Uncompute $B$ and $A$.
\end{enumerate}
}

The algorithm for checking whether any of the clauses is violated is given as Algorithm \ref{alg:psat}. Checking whether $x$ violates an individual clause can be achieved with just one Toffoli gate, controlled on $k$ bits (together with some additional NOT gates). This is because we use 01, 10 to represent a variable being set to false and true, respectively; so controlling on one or other of those bits is enough to check for an assignment being consistent with a literal in a clause.

Step 2 of the algorithm checks whether any of the bits of $x$ are equal to the fictitious value represented by the binary string 11. This can be achieved using an $n$-qubit ancilla register $y$, such that bit $y_i$ is set to 1 if both of these bits are equal to 1, using a Toffoli gate. Then $A_{m+1}$ is set to 1 if any of the bits $y_i$ are equal to 1, and then $y$ is uncomputed.

Then steps 4 and 5 of the algorithm determine whether the partial assignment $x$ is complete. If the assignment is complete, and not inconsistent with any of the clauses, the output is set to true.


\subsection{Evaluation of $h(x)$: choosing the next variable}
\label{sec:hsat}

Given our simple choice of heuristic for $k$-SAT, the $h$ function is also very simple: it returns $\ell + 1$, which can be achieved via copying $\ell$ and then incrementing it in place.


\section{Classical graph colouring algorithms}
\label{sec:dsatur}

Graph colouring algorithms can be categorised as either heuristic or exact. Heuristic algorithms aim to output a ``good'' colouring (one with a low number of colours) without proving optimality, while exact algorithms allow the impossibility of colouring with a certain number of colours to be certified. In this work the focus is on exact algorithms, corresponding to problems for which it is crucial to determine optimality. For example, a wireless communication problem where it is extremely expensive to use unnecessarily many frequencies (we might imagine that there is a fixed allocation of bandwidth across all cells in a wireless communication system, so having more colours would correspond to each cell having lower bandwidth). A vast array of heuristic algorithms for graph colouring has been proposed; for a relatively recent survey of these (and exact algorithms), see~\cite{malaguti10}.

The idea of choosing a vertex to colour that is the most ``saturated'' was proposed by Br\'elaz in 1979~\cite{brelaz79} as a heuristic, called DSATUR, for finding a good colouring. This idea can also be used as part of a backtracking algorithm, also called DSATUR. The full algorithm, incorporating an optimisation due to Sewell~\cite{sewell96}, can be summarised as follows:

\begin{enumerate}
\item Find a large (though possibly non-maximal) clique in the graph via an efficient algorithm, and colour the vertices in that clique.
\item Perform the standard backtracking algorithm with the following heuristic $h$ for choosing the next vertex to colour:
\begin{enumerate}
\item Choose the vertex which is adjacent to the largest number of vertices with distinct colours.
\item In case of a tie in (a), choose the vertex with the largest degree.
\item In case of a tie in (b), choose the lexicographically first vertex.
\end{enumerate}
The $P$ function is defined to reject any partial colouring that is invalid.
\end{enumerate}

There are two simple further optimisations that are implemented in standard versions of DSATUR~\cite{mehrotra96}. The first is, when the current partial colouring contains colours between 1 and $c$, to only consider colours between 1 and $c+1$ for colouring the next vertex. The second is not to assign a vertex a colour that is already used by any of its neighbours. The former optimisation can be implemented in the quantum backtracking algorithm quite efficiently (see above), whereas the latter seems less straightforward to achieve without incurring a multiplicative loss of $O(\sqrt{k})$ in the complexity.

DSATUR can also be used to compute the chromatic number, rather than checking $k$-colourability. In this (more standard) variant, the number of colours used can be as large as $n$. When a complete and valid colouring of the graph is found that uses fewer colours than the best colouring found so far, the number of colours it uses is stored and any further partial colourings that use more than that number of colours are rejected. Then the algorithm finally outputs the stored (minimal) number of colours. 

The clique-finding step, which affects the runtime very little, can be implemented classically before the rest of the algorithm is executed on the quantum computer. The tie-breaking steps can be built into the $h$ function presented in Section \ref{sec:h} with little extra effort, by sorting the vertices according to degree before running the algorithm.

The DSATUR algorithm is very simple, but remains a competitive approach for colouring random graphs, called a ``de facto standard''~\cite{sewell96}, performing ``suprisingly well'' and used on small subproblems as part of other algorithms~\cite{sansegundo12}. (For large structured graphs, approaches based on expressing the colouring problem as an integer program and solving this via relaxations seem to be superior~\cite{malaguti10}.) Further, a standard implementation is available\footnote{\url{http://mat.gsia.cmu.edu/COLOR/solvers/trick.c}}, and this algorithm is widely used as a benchmark in colouring competitions.

Additional modifications to the DSATUR algorithm were suggested by Sewell~\cite{sewell96} and San Segundo~\cite{sansegundo12} which achieve improved runtimes for certain graphs (e.g.\ up to about a factor of 2 in results reported in~\cite{sansegundo12}). Mehrotra and Trick presented a new linear programming (LP) relaxation and compared it against DSATUR~\cite{mehrotra96}. Results were mixed; on some random instances the LP relaxation substantially outperformed DSATUR, and on others the reverse was true. For structured instances, the LP approach was often (though not always) superior. Mixed results are also seen in the comparison given in~\cite{mendezdiaz06} of DSATUR against a new branch-and-cut method: the new method (which also uses DSATUR as a subroutine) usually, but not always, outperforms DSATUR itself. By contrast, in the experimental results reported in~\cite{sansegundo12}, DSATUR always outperforms a competitor branch-and-price algorithm on random instances. Interestingly, despite significant progress on the development of fast SAT solvers over recent years, the approach of encoding a graph colouring as a SAT instance and solving it using a SAT solver does not seem to be particularly effective for random graphs~\cite{vangelder08}.

Given all the above results, together with the algorithm's ubiquity, we believe that DSATUR is a reasonable choice of classical algorithm for benchmarking purposes.


\section{Classical experimental results}
\label{sec:classexp}

In order to determine the likely performance on large instances of the classical algorithms considered, and to calculate the quantum backtracking algorithm's likely performance, we evaluated the classical algorithms on many random instances to determine their runtime scaling with problem size. We now describe the results of these experiments.


\subsection{Satisfiability}

\begin{table}
\begin{center}
\begin{tabular}{|c|c|c|c|}
\hline $k$ & $\alpha_k$ & \fontfamily{cmss}\selectfont Exponent & $n_{\max}$ \\
\hline 3 & 4.27 & $0.03n - 4.88$ & 425\\
4 & 9.93 & $0.11n - 7.10$ & 132\\
5 & 21.12 & $0.20n - 7.85$ & 84\\
6 & 43.37 & $0.23n - 7.38$ & 60\\
7 & 87.79 & $0.34n - 9.51$ & 54\\
8 & 176.54 & $0.42n - 11.12$ & 47\\
9 & 354.01 & $0.55n -13.51$ & 41\\
10 & 708.92 & $0.55n -12.96$ & 41\\
11 & 1418.71 & $0.55n - 12.00$ & 36\\
12 & 2838.28 & $0.56n -10.86$ & 37\\
13 & 5677.41 & $0.55n - 9.39$ & 36\\
14 & 11355.67 & $0.51n - 6.55$ & 33\\
15 & 22712.20 & $0.46n - 4.38$ & 31\\
\hline
\end{tabular}
\end{center}
\caption{Estimated runtime scaling with $n$ of the Maple\_LCM\_Dist SAT solver on random $k$-SAT instances with $n$ variables and $\approx \alpha_k n$ clauses. Table lists suspected exponents $f(n)$ in runtimes of the form $2^{f(n)}$, measured in CPU-seconds. Based on taking linear least-squares fits to the log of median runtimes from $\ge 100$ random instances, omitting instances where $n < 100$ for $k=3$; $n < 60$ for $k=4$; $n < 50$ for $k=5$. $n_{\max}$ column lists the maximal value of $n$ considered in each case.}
\label{tab:satexponents}
\end{table}

We ran the Maple\_LCM\_Dist SAT solver, the winner of the ``main track'' of the SAT Competition 2017~\cite{balyo17}, on randomly generated instances of $k$-SAT, for various choices of $k$. 
Experiments were run on an Intel Core i7-4790S CPU operating at 3.20GHz with 7GB RAM. We generated the random instances using CNFgen, a standard tool which picks $k$-SAT instances according to the distribution described in Section \ref{sec:results}. For each $k$ and $n$, we chose a number of clauses $m$ such that $m \approx \alpha_k n$, where $\alpha_k$ is the threshold for $k$-SAT. The value of the threshold is not known precisely for all $k$. Indeed, it was only recently shown rigorously that a sharp threshold exists for large $k$~\cite{ding15}; for small $k \ge 3$ this is still unknown. The tightest bound known for general $k$ is that
\be \label{eq:threshold} 2^k \ln 2 - \frac{3 \ln 2}{2} - o(1) \le \alpha_k \le 2^k \ln 2 - \frac{1 + \ln 2}{2} + o(1), \ee
where the $o(1)$ terms approach 0 as $k \rightarrow \infty$~\cite{cojaoghlan13}. Non-rigorous predictions for the threshold in the range $3 \le k \le 7$ obtained via the ``cavity method'' are given in~\cite{mertens06}. Here we used these predictions as our estimates for $\alpha_k$ for $k \le 7$. In the range $8 \le k \le 10$, we used the (non-rigorous) upper bounds $\alpha_c^{(7)}$ given in~\cite[Table 1]{mertens06}, while an approximate threshold in the range $k > 10$ can be found via the third-order approximation given in~\cite[Appendix]{mertens06}. For large $k$, this approximation rapidly approaches the upper bound in (\ref{eq:threshold}). This method does not guarantee that we have found a good approximation to the true threshold $\alpha_k$; however, the level of accuracy of this approximation seems to be sufficient for the small input sizes that we were able to consider. It could also be possible that another choice of $\alpha_k$ away from the threshold could produce even harder instances for the Maple\_LCM\_Dist solver.

\begin{figure}
\includegraphics[width=\columnwidth]{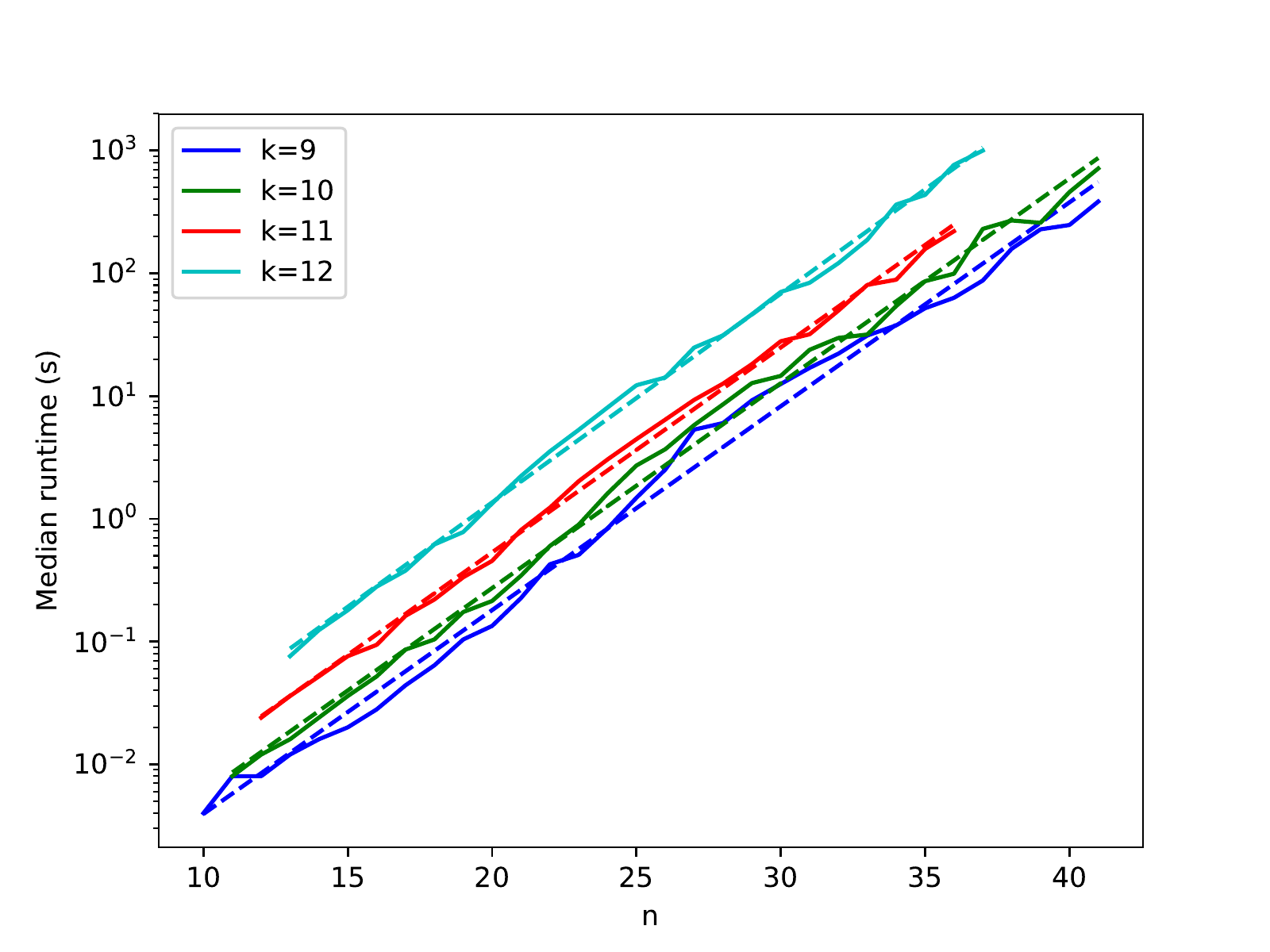}
\caption{Runtime of the Maple\_LCM\_Dist SAT solver on random $k$-SAT instances with $n$ variables and $\approx \alpha_k n$ clauses. Solid line represents the median of at least 100 runs, in CPU-seconds. Dashed lines are linear least-squares fits.}
\label{fig:satscaling}
\end{figure}

\begin{figure}
\includegraphics[width=\columnwidth]{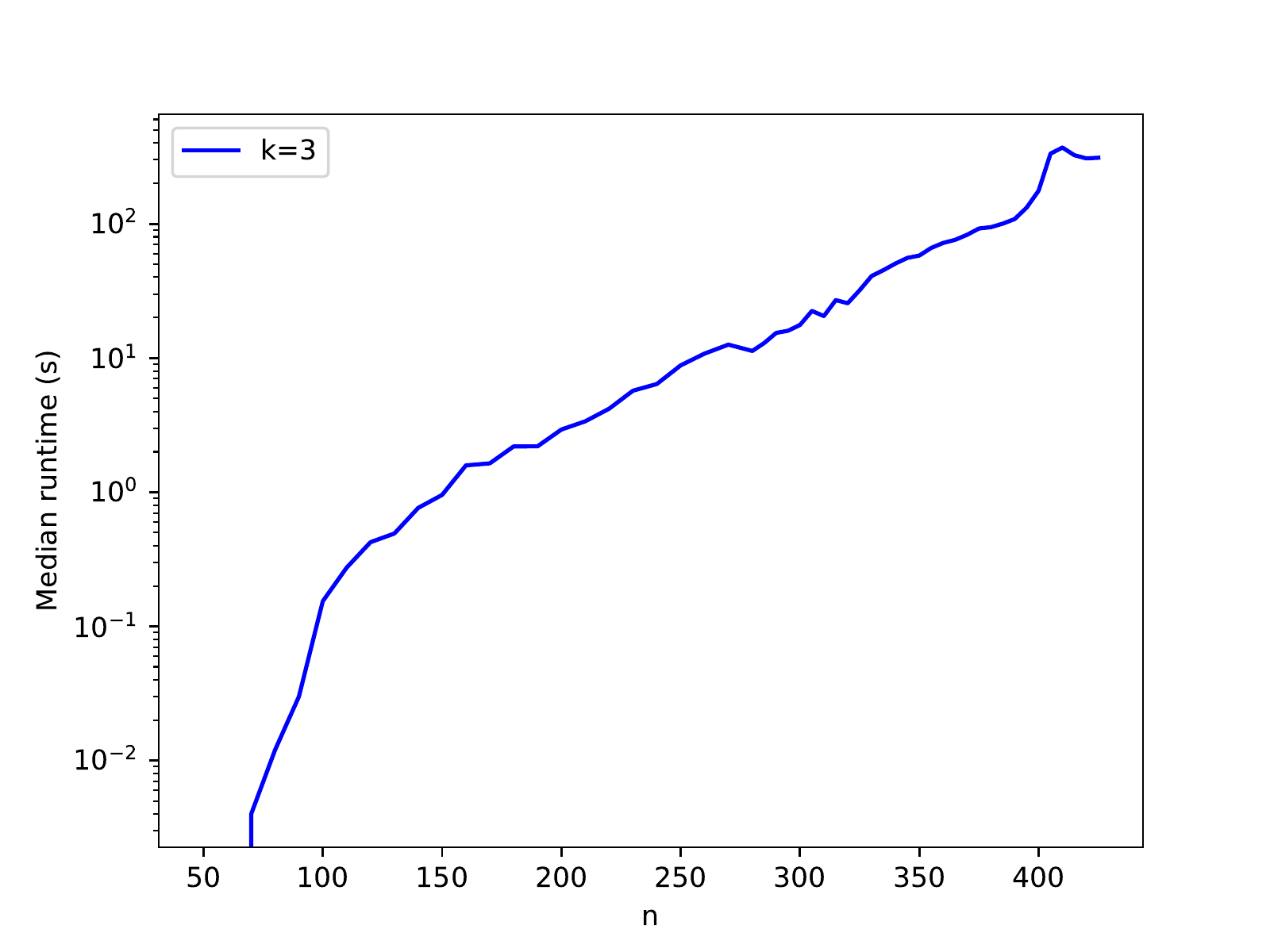}
\caption{Median runtime, in CPU-seconds, of 100 runs of the Maple\_LCM\_Dist SAT solver on random 3-SAT instances with $n$ variables and $\approx 4.267n$ clauses.}
\label{fig:3satscaling}
\end{figure}

For each value of $n$ considered, we ran the solver on at least 100 random instances and took the median runtime. As we choose $m$ approximately at the threshold value for satisfiability, the random instances generated may be either satisfiable or unsatisfiable, and we did not filter them by (e.g.)\ only choosing satisfiable instances. The estimated scaling parameters for each $k$ are listed in Table \ref{tab:satexponents}, and examples of the results for $k \in \{9,\dots,12\}$ are shown in Figure \ref{fig:satscaling}. Note that, given that we only consider relatively small values of $n$, we cannot rule out that the runtime does not simply scale as a function $f(n) = 2^{a n + b}$ for some $a,b \in \R$. For example, in the case $k=3$, the runtime behaviour seems to be more complex in the range that we considered (see Figure \ref{fig:3satscaling}).

\begin{table}
\begin{center}
\begin{tabular}{|c|c|c|c|}
\hline $k$ & $\alpha_k$ & \fontfamily{cmss}\selectfont Exponent\\
\hline 3 & 4.267 & $0.35n + 3.70$ \\
4 & 9.931 & $0.46n + 3.65$ \\
5 & 21.117 & $0.54n + 3.52$ \\
6 & 43.37 & $0.60n + 3.46$ \\
7 & 87.79 & $0.64n + 3.45$ \\
8 & 176.54 & $0.68n + 3.43$ \\
9 & 354.01 & $0.70n + 3.46$ \\
10 & 708.92 & $0.72n + 3.52$\\
11 & 1418.71 & $0.74n + 3.61$ \\
12 & 2838.28 & $0.75n + 3.68$ \\
13 & 5677.41 & $0.76n + 3.81$ \\
14 & 11355.67 & $0.78n + 3.72$ \\
15 & 22712.20 & $0.79n + 3.76$ \\
\hline
\end{tabular}
\end{center}
\caption{Estimated backtracking tree size on random $k$-SAT instances with $n$ variables and $\approx \alpha_k n$ clauses, when variables are ordered in terms of appearance count. Table lists suspected exponents $f(n)$ in tree sizes of the form $2^{f(n)}$. Based on taking linear least-squares fits to the log of median runtimes, from: 15 random instances and $10 \le n \le 30$ for $k \le 9$; 7 random instances and $15 \le n \le 25$ for $10 \le k \le 12$; 5 random instances and $15 \le n \le 20$ for $13 \le k \le 15$ (omitting $n=15$ in the case $k=15$).}
\label{tab:satbtexponents}
\end{table}

\begin{figure}
\includegraphics[width=\columnwidth]{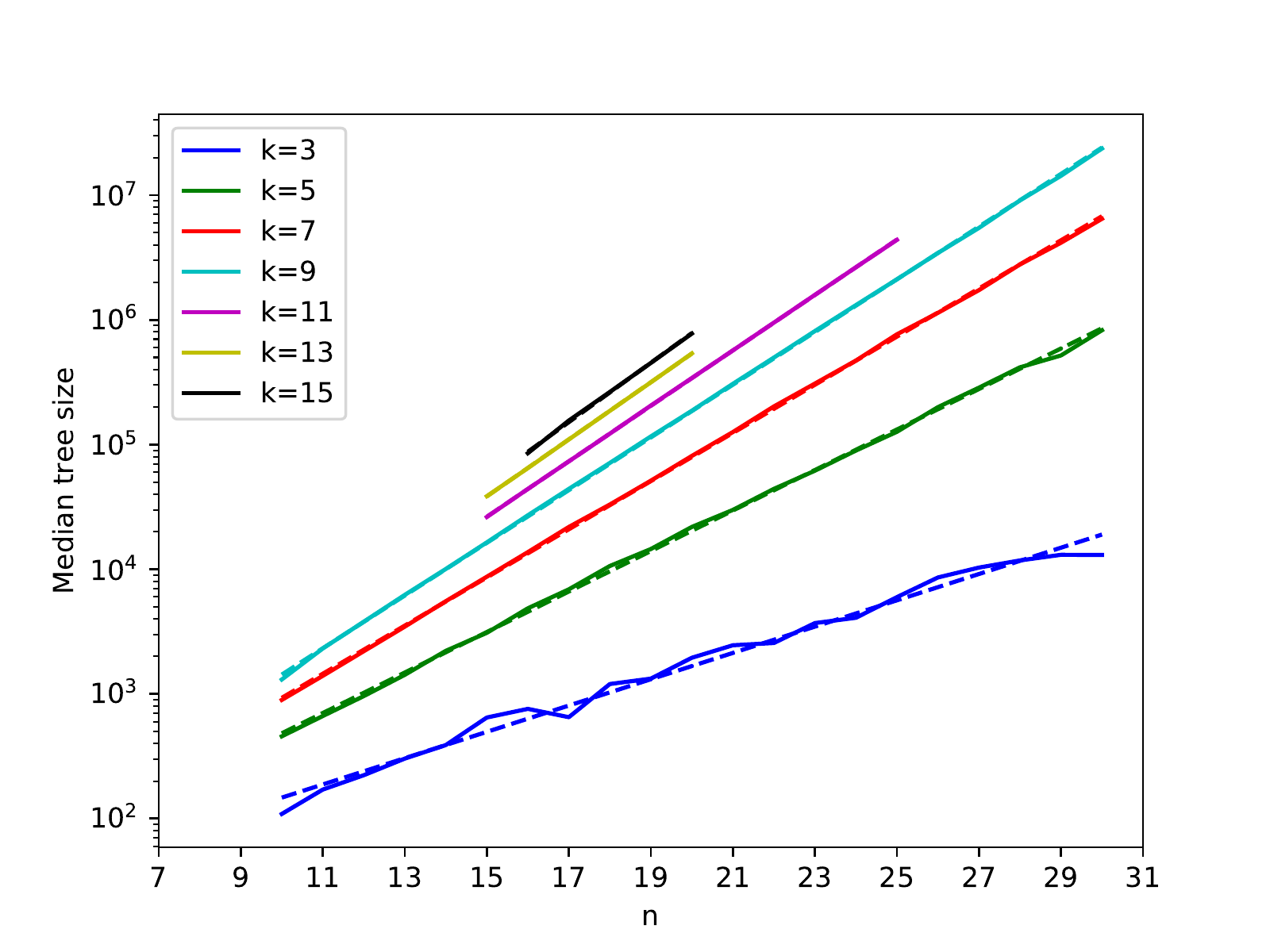}
\caption{Backtracking tree size when variables are ordered in terms of appearance count. Solid line represents the median on random $k$-SAT instances; number of runs as described in Table \ref{tab:satbtexponents}. Dashed lines are least-squares fits.}
\label{fig:satbtscaling}
\end{figure}

In the case of backtracking, we evaluated the size of the backtracking tree produced for the standard backtracking algorithm applied to random $k$-SAT instances where the heuristic $h$ chooses the next variable from a variable ordering fixed in advance, in order of number of appearances (that is, the variable that appears in the largest number of clauses is chosen first). Some previous comparisons of static variable ordering strategies for solving CSPs have been made in~\cite{dechter89,balafoutis10}. A good static ordering strategy can produce substantially smaller backtracking trees than choosing the variables in a fixed order that does not depend on the problem instance (see~\cite{brown81,montanaro18} for analytic expressions for the expected size of the latter).


\subsection{Graph colouring}
\label{sec:classexpgc}

\begin{figure}[t]
\includegraphics[width=\columnwidth]{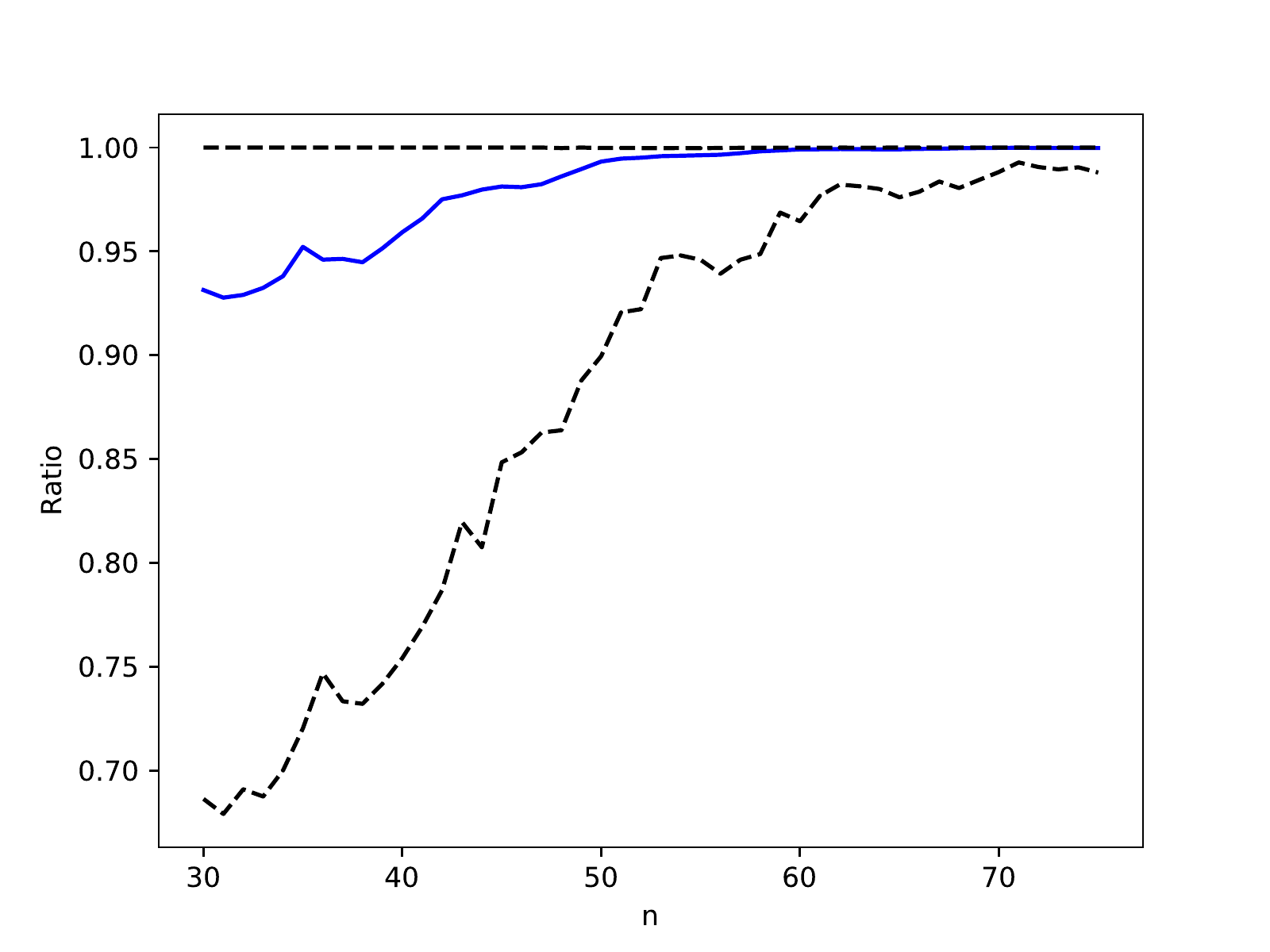}
\caption{Ratios between the number of nodes in the DSATUR backtracking trees for computing chromatic number and checking $k$-colourability, over 1000 random graphs for each $n \in \{30,\dots,75\}$. Solid line: median; dashed lines: 5th/95th percentiles.}
\label{fig:colchromaticratio}
\end{figure}

We experimentally evaluated the DSATUR algorithm for 1000 random graphs in the range $n \in \{10,\dots,75\}$, where each edge is present with probability 0.5. As discussed in Section \ref{sec:dsatur}, the algorithm can either be used to calculate the chromatic number precisely, or to determine $k$-colourability for a fixed $k$ (which matches what the quantum algorithm achieves and is, in principle, easier\footnote{Where ``easier'' should be interpreted from a practical point of view; from the perspective of computational complexity theory, the two problems are polynomial-time equivalent.}). The second variant can be obtained from the first by rejecting any colouring which uses more than $k$ colours. In order to determine a challenging value of $k$ for a given graph $G$, we first run the first variant of DSATUR to calculate the chromatic number $\chi(G)$, then the second variant with $k=\chi(G)$. In Figure \ref{fig:colchromaticratio}, we verify that for most random graphs, $\chi(G)$-colouring is not significantly easier than computing $\chi(G)$.

\begin{figure}[t]
\includegraphics[width=\columnwidth]{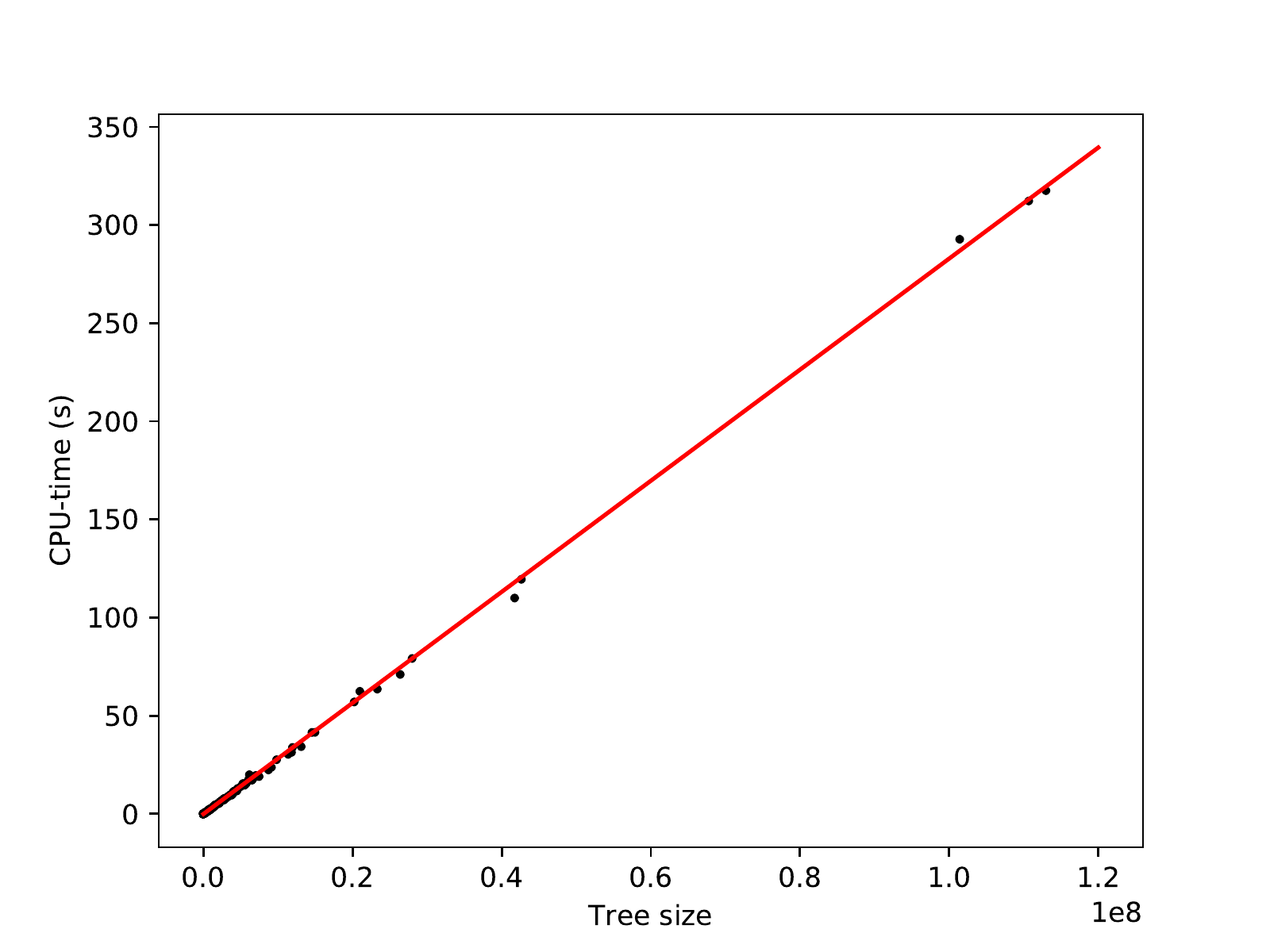}
\caption{Number of nodes in the DSATUR $k$-colourability backtracking tree vs.\ runtime in CPU-seconds on a 3.20GHz Intel Core i7-4790S processor. Scatter plot showing 1\% of the 1000 random graphs considered for each $n \in \{10,\dots,75\}$. Line is a least-squares fit.}
\label{fig:treevstime}
\end{figure}

We also observe that, as expected, the number of nodes $T$ in the backtracking tree has a strong linear correlation with the CPU time $\tau$ used (see Figure \ref{fig:treevstime}). Performing a least-squares fit, we obtain that on a 3.20GHz Intel Core i7-4790S processor, $\tau \approx 2.50\times 10^{-6} T - 0.05$, where $\tau$ is measured in seconds. This corresponds to each node in the backtracking tree being evaluated in $\approx 8000$ CPU cycles. Although the number of operations per node in the backtracking tree does depend on $n$, the dependence is linear, so the scaling should remain correct up to a factor of 3 or so for all reasonable graph sizes. Therefore, $T$ is a good proxy for the actual runtime of the algorithm.

\begin{figure}[t]
\includegraphics[width=\columnwidth]{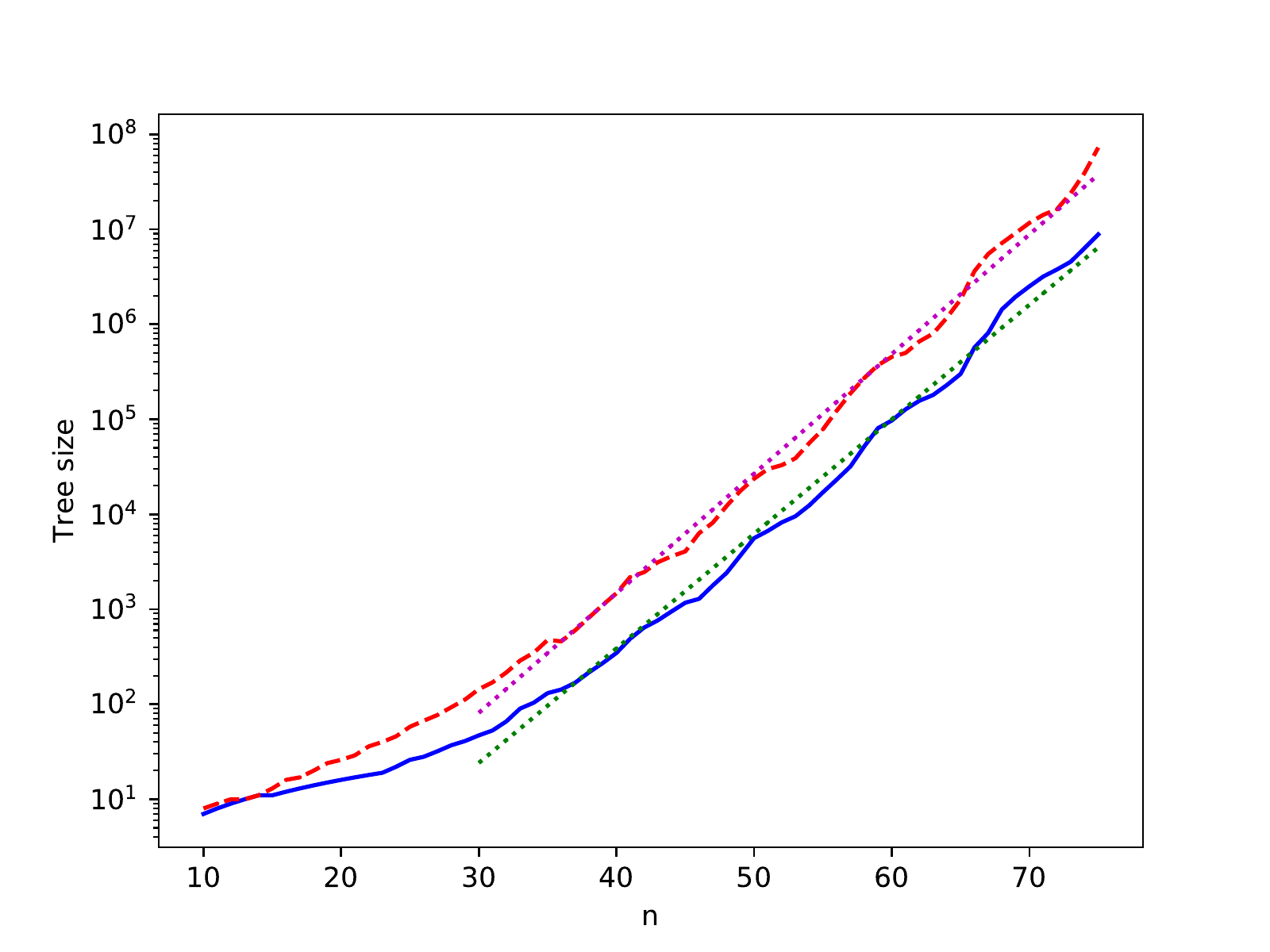}
\caption{Number of nodes in the DSATUR $k$-colourability backtracking tree. Median (solid) and 90th percentile (dashed) over 1000 random graphs for each $n \in \{10,\dots,75\}$. Dotted lines are least-squares fits for the range $n \ge 30$.}
\label{fig:nvstreesize}
\end{figure}

Next, we calculate how $T$ scales with $n$. We compute both the median value and the 90th percentile, where the latter aims to provide an estimate for how the runtime will scale for the most difficult graphs. As expected, these quantities scale exponentially with $n$. Performing a least-squares fit on $\log T$ (omitting small values of $n$), we obtain that $T \approx 2^{0.40n -7.43}$ for the median, and $T \approx 2^{0.42n -6.20}$ for the 90th percentile; see Figure \ref{fig:nvstreesize}. Throughout this work, we use the median to compute the anticipated runtime of classical and quantum algorithms on random graphs.

The quantum algorithm presented here can be seen as accelerating a somewhat simplified version of DSATUR, without an optimisation to rule out colours that are used by each vertex's neighbours. So, when computing an estimate for the backtracking tree size of this simplified algorithm (to obtain a corresponding estimate of the quantum algorithm's runtime), we should take into account the cost of not including this optimisation. This cost will vary depending on the input. Further, DSATUR has two roles: for a graph with chromatic number $k$, it both finds a $k$-colouring, and rules out any $(k-1)$-colouring. The quantum algorithm, by contrast, only determines colourability. When comparing the two algorithms, we therefore measure DSATUR searching for a $k$-colouring against the simplified algorithm ruling out a $(k-1)$-colouring.

\begin{figure}
\includegraphics[width=\columnwidth]{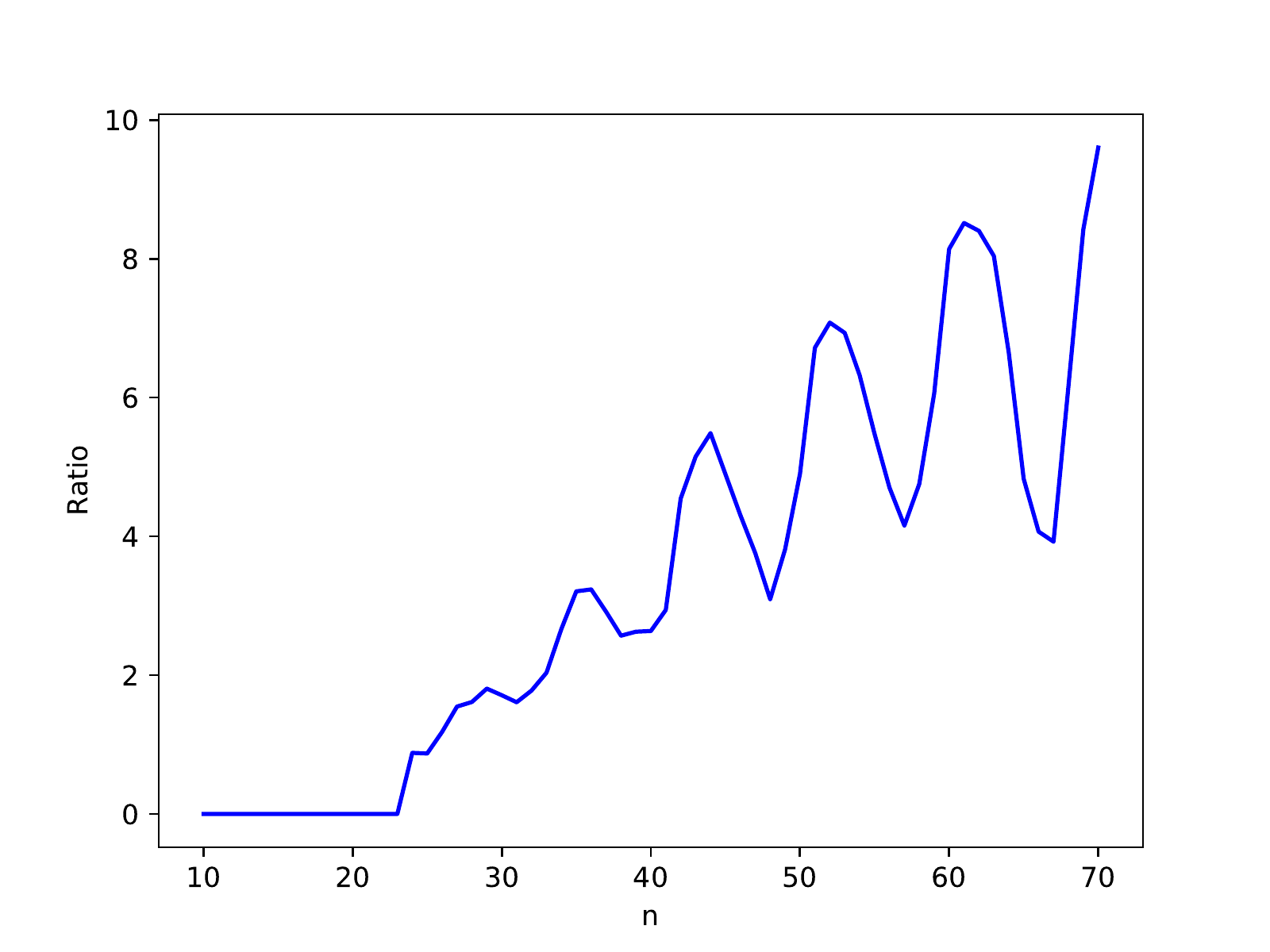}
\caption{Median ratio between the number of nodes in the unoptimised DSATUR backtracking tree, as used in the quantum algorithm, and the number of nodes in the optimised DSATUR backtracking tree. Taken over 1000 random graphs for each $n \in \{10,\dots,75\}$.}
\label{fig:ratio-bt-trick}
\end{figure}

We expect the ratio of the tree sizes of the simplified algorithm and the standard algorithm to usually be relatively small (though perhaps growing slowly with $n$); it is always upper-bounded by the number of colours. In some cases, the simplified algorithm may have {\em smaller} tree size, because it aims only to prove that a $(k-1)$-colouring does not exist. This effect is particularly significant in the case where the clique-finding preprocessing step finds a $k$-clique, enabling a $(k-1)$-colouring to be ruled out immediately; this event is quite common for small random graphs. Experimentally, the median ratio of tree sizes varies for different choices of $n$; in all experiments run, the ratio was less than 15. The median ratio found for each $n$ is plotted in Figure \ref{fig:ratio-bt-trick}. It is apparent that this varies and, for some fairly large values of $n$, fluctuates below 4. As we aim to find a ``best case'' but fair scenario for the quantum algorithm outperforming its classical counterpart, we use a factor of 4 as our estimate of the penalty to the quantum algorithm's backtracking tree size. Put another way, we are assuming that the algorithms are run on random graphs for a ``good'' choice of $n$.

\section{Cost of classical computation}
\label{sec:classcost}

Under some models, the quantum algorithms we consider seem to achieve a substantial speedup over their classical competitors. However, the depth-optimised algorithms we describe use substantial physical quantum resources. In order to determine if we obtain a real-world reduction in cost, we need to compare the cost of classical and quantum computation. Given that early quantum computers are likely to be accessed remotely as a cloud service, perhaps a fair model for comparison is to consider a cloud-based compute service which charges by hour of CPU time. Then one can calculate the number of CPU-hours required to solve the instance which can be solved by the quantum computer in 1 day, allowing a price to be put on ``1 quantum CPU-hour''. See~\cite{markov18} for a very recent example of a similar comparison in the context of quantum computational supremacy experiments. The classical algorithm used may not be perfectly parallelisable across multiple machines; however, for the algorithms that are typically used to solve hard constraint satisfaction problems in practice (e.g.\ backtracking, local search), a relatively high level of parallelisation may be possible~\cite{rao93}. 

In order to make a fair cost comparison between the quantum and classical machines, it is also necessary to take into account the cost of the classical processing used in the quantum computer. For example, if each physical qubit requires one classical CPU to carry out the error-correction computations for the surface code, there are $10^6$ qubits, and the computation takes 1 day, we could instead use these classical CPUs to carry out a computation requiring $10^6$ CPU-days. We therefore calculate the overall cost of the quantum computation in terms of the {\em classical} CPU time\footnote{We would like to thank Craig Gidney for suggesting this cost model.}, and use the ratio between this and the classical CPU time used as the quantum speedup factor. Although this measure is a lower bound on the true cost of the quantum computation (which should also take into account issues such as power consumption), it is independent of the details of the quantum hardware platform itself.

The classical processing required may not be carried out using a standard CPU, but instead using a GPU or specialised electronics (FPGAs or ASICs). The extent of the speedup that might be possible by this approach can be estimated using Bitcoin mining as a point of comparison. Current commercially available hardware platforms based on ASICs can achieve a hash rate of over $2 \times 10^{13}$ hash function evaluations per second\footnote{See \url{https://en.bitcoin.it/wiki/Mining_hardware_comparison}, for example. Note that one ``hash'' corresponds to two SHA-256 hash function evaluations.}, as compared with a rate of around $2 \times 10^9$ for a GPU or $10^7$ for a CPU. The best average runtime reported for a recent near-linear time decoding algorithm for the surface code~\cite{delfosse17}, using a standard CPU, is approximately 440$\mu$s for a system of 5000 qubits, corresponding to $8.8 \times 10^{-8}$s per qubit. (Note that this latter figure is likely to be an overoptimistic estimate, as this would imply that decoding the surface code could be perfectly parallelised; on the other hand, algorithmic and implementation optimisations to the algorithm of~\cite{delfosse17} could improve its runtime.)

So one CPU could support around 2 qubits while still achieving a surface code cycle time of 200ns.
Assuming that GPU/ASIC implementations of this decoding algorithm are indeed possible with roughly similar performance enhancements to the case of Bitcoin, we might hope that a GPU-based system could support 100 times as many qubits at this clock speed, and that an ASIC could support $10^6$ times as many qubits. (Increased clock speeds would require a corresponding increase in classical performance.) In unpublished work~\cite{fowlerpc}, it has been estimated that a CPU core and FPGA could realistically perform error correction for around 100 physical qubits, which would be 1--2 orders of magnitude faster than a CPU alone; so this is comparable to what we here call a GPU.

An indication of the computational resources required to implement different numbers $N$ of Toffoli gates is shown in Table \ref{tab:proccost} (the figures would be similar for T gates). This table is calculated based on the cost of implementing Toffoli factories alone. First, the number of qubits required for each Toffoli factory is estimated. Then the processing time required by the decoder to support each Toffoli factory is determined by multiplying the estimated time for decoding each qubit (based on~\cite{delfosse17}, scaled according to processor type) by the spacetime cost of the Toffoli factory. Finally, this is multiplied by the number of Toffoli gates to get an estimate of overall processing cost.

This is clearly a very approximate calculation, yet may give an indication of the effect of this classical overhead. (For example, error-correction procedures may run more quickly if there are fewer errors, and we have not considered this effect here.) By comparing Tables \ref{tab:speedups} and \ref{tab:proccost}, one can see that under all of the regimes, and even if ASICs are used, the complexity of classical processing wipes out any significant advantage for the quantum algorithms. One reason for this is that the Toffoli counts of the algorithms are substantially higher than the T-depths. This may motivate (especially in the Optimistic scenario) the use of an alternative error-correction scheme with lower overhead, albeit perhaps a worse threshold than the surface code.

\begin{table}
\fontfamily{cmss}\selectfont
\begin{center}
\begin{tabular}{|c|c|c|c|}
\hline $N$ & Realistic & Plausible & Optimistic \\
\hline $10^{12}$ & $4.17 \times 10^{7}$ & $4.30 \times 10^{4}$ & $9.15 \times 10^{-1}$\\
\hline $10^{16}$ & $2.29 \times 10^{12}$ & $7.76 \times 10^{8}$ & $2.23 \times 10^{4}$\\
\hline $10^{20}$ & $3.10 \times 10^{16}$ & $3.07 \times 10^{13}$ & $3.28 \times 10^{8}$\\
\hline
\end{tabular}
\end{center}
\caption{Classical processing required to implement $N$ Toffoli gates under different regimes, based on extrapolation of runtimes reported in~\cite{delfosse17}. Measured in processor-days (where type of processor is CPU, GPU and ASIC respectively in realistic, plausible and optimistic regimes). Assumes that the speedup offered by GPUs and ASICs over CPUs is a factor of 100 and $10^6$ respectively.}
\label{tab:proccost}
\end{table}


\section{Conclusions and further work}
\label{sec:conclusions}

For the first time, we have given a detailed analysis of the complexity of quantum algorithms for graph colouring and boolean satisfiability, including overheads for fault-tolerance, and have shown that in some scenarios the algorithm could substantially outperform leading classical competitors in terms of runtime. However, when one takes into account the cost of classical processing using current techniques, the speedup disappears. Also, the space usage of the algorithms is extremely high (sometimes over $10^{13}$ physical qubits), although this could be reduced at the expense of a longer runtime, by changing the algorithms to perform fewer tasks in parallel. Simply allowing the algorithms to run for longer will increase the size of the potential speedup too; for example, allowing the backtracking algorithm for $k$-colouring to run for $c$ days would increase the speedup by a factor of approximately $c$.

There are some theoretical improvements that could be considered to improve the complexity of these algorithms. Arunachalam and de Wolf have given a variant of Grover's algorithm which solves the unstructured search problem for a unique marked element in a set of $N$ elements using only $O(\log(\log^* N))$ gates between each oracle query~\cite{arunachalam17}, where $\log^* N$ is the number of times the binary logarithm function must be applied to $N$ to obtain a number that is at most 1. It might be possible to use these ideas to accelerate the variant of Grover's algorithm that we used here. (Note that the {\em depth} of the circuit between each oracle query in the variant of the algorithm that we use is only of size $O(\log \log N)$ already, so the gain might be relatively minor.)

Some specific ways in which it might be possible to improve the quantum backtracking algorithm are:

\begin{itemize}
\item The main component of the algorithm is a controlled quantum walk operation (controlled-$R_B R_A$) used within phase estimation. The use of controlled gates throughout adds some overhead to the algorithm. If it were possible to just run the quantum walk directly, rather than needing to apply phase estimation to it, this would reduce the complexity (and could also give a more efficient algorithm for finding a solution, rather than detecting its existence). Preliminary calculations suggest that replacing controlled operations with uncontrolled ones could reduce the runtime by up to $\sim$20\%.

\item An automated circuit optimisation tool, such as T-par~\cite{amy13}, could be used to reduce the complexity of the quantum circuits developed, and to check correctness. It is not obvious how large a reduction could be achieved, given that some of the subroutines used (such as integer addition) are already low-depth and highly optimised.

\item One could consider optimisations targeted at particular classes of graphs (e.g.\ sparse graphs) or boolean formulae, and could consider different heuristic functions $h$.
\end{itemize}

Using the quantum backtracking algorithm to actually find a $k$-colouring when one exists, rather than simply determining whether one exists, would increase its complexity. However, careful use and optimisation of the techniques of~\cite{ambainis17,jarret18} could minimise this overhead.

In this work, the backtracking algorithm has been aggressively optimised for circuit depth. It is unclear how much further this process can be continued. Given that the depth overhead of the quantum circuit is within an order of magnitude of the quantum circuit for SHA-256~\cite{amy16}, it seems plausible that one or two further orders of magnitude improvement are the most that will be possible. However, it may be possible to reduce the Toffoli-count of the algorithm, which is extremely high at present. Also, even without considering the overhead for fault-tolerance, the space usage of the algorithm in general seems very large ($\gg 10^5$ logical qubits in the case of graph colouring), although it is worth bearing in mind that the input size itself is relatively high (e.g.\ to describe an arbitrary graph on 150 vertices requires over $10^4$ bits). A further issue faced by the quantum backtracking algorithm is its inability to retain state across different evaluations of oracle functions, whereas this is available to the classical algorithm and can make it more efficient.

Our results suggest that improved fault-tolerance techniques will be required to make the algorithms presented here truly practical. (See~\cite{litinski18,fowler18} for some very recent developments in this direction.) In particular, if a significant quantum speedup is to be realised, it seems to be essential to have highly efficient decoding algorithms and/or specialised decoding hardware. Although the numbers reported here are daunting, there is scope for improvement. A potentially hopeful parallel is previous work on quantum algorithms for quantum chemistry, where initially reported complexities were high~\cite{wecker14}, but were rapidly improved by orders of magnitude (see~\cite{babbush18} and references therein). This was achieved by a careful analysis of optimisations to the specific algorithms used, and numerical calculations to determine performance of the algorithms in practice, beyond the theoretical worst-case bounds. Each of these general strategies could be applied here.

The analysis here could be extended by attempting to find more realistic models for constraint satisfaction problems that occur in practice, rather than the completely unstructured families considered here. However, these can be seen as representing the ``core'' of a challenging problem, so are perhaps reasonable in themselves. A more detailed comparison could also be made with other classical solvers, including the use of special-purpose hardware~\cite{sohanghpurwala17}.


\subsection*{Acknowledgements}

This work was supported by EPSRC grants EP/L021005/1, EP/L015730/1, EP/M024261/1 and EP/R043957/1, and EPSRC/InnovateUK grant EP/R020426/1. We acknowledge support from the QuantERA ERA-NET Cofund in Quantum Technologies implemented within the European Union's Horizon 2020 Programme (QuantAlgo project). We would like to thank many people including Steve Brierley, Austin Fowler, Craig Gidney, Naomi Nickerson, Stephen Piddock and Cathy White, as well as the rest of the QCAPS project team, for helpful discussions and comments relating to this work. Supporting data are available at the University of Bristol data repository, data.bris, at \url{https://doi.org/10.5523/bris.17mjvdv1u7udm2u9frlh61ltvp}.

\vspace{10pt}


\appendix

\section{Resource requirements for Toffoli state factories}
\label{app:tofffactory}

Here we include, as Algorithms \ref{alg:tofffactory} and \ref{alg:tfactory}, algorithms for computing the resource requirements for Toffoli and T state factories, based on the analysis in~\cite{ogorman17} and presented similarly to the T state factory algorithm in~\cite{aggarwal17}.  Some approximations are made to present a simple algorithm, but in the limit $p_g \leq 10^{-3}$ these are extremely accurate.

\boxalgm{alg:tofffactory}{Resource requirements for a Toffoli factory}{
{\bf Input:} gate error $p_g$, number $N$ of Toffoli gates in circuit, $t_{\mathrm{algo}}$ time of algorithm in units of surface code cycles;
\\
{\bf Output:} $Q$ number of physical qubits used by factory.
\begin{enumerate}
\item $p_{\operatorname{tol}} \leftarrow 1/(3N)$
\item $i \leftarrow 1$
\item $d_{i} \leftarrow \min \{ d \in \N: 99 d (100 p_g)^{(d+1)/2} \ge p_{tol} \}$
\item $Q_{i} \leftarrow 44 d_i(d_i-1)$
\item $T_{i} \leftarrow 9 d_i$
\item $p_{\operatorname{tol}} \leftarrow \sqrt{p_{\operatorname{tol}}/28}$
\item While $p_{\operatorname{tol}} < p_g$:
\begin{enumerate}
\item $i \leftarrow i+1$
\item $d_i \leftarrow \min \{ d \in \N: 250 d (100 p_g)^{(d+1)/2} \ge p_{\operatorname{tol}} \}$
\item $Q_i \leftarrow (8*15^{i-2})100 d_i(d_i-1)$
\item $T_i \leftarrow 10 d_i$
\item $p_{\operatorname{tol}} \leftarrow (p_{\operatorname{tol}}/36  )^{1/3}$
\end{enumerate}
\item layers $\leftarrow i$
\item $S=\sum_{j=1}^{\mathrm{layers}} Q_j T_j$
\item  Return $Q=N S / t_{\mathrm{algo}}$. 
\end{enumerate}
}

\boxalgm{alg:tfactory}{Resource requirements for a T state factory}{
	{\bf Input:} gate error $p_g$, number $N$ of T gates in circuit, $t_{\mathrm{algo}}$ time of algorithm in units of surface code cycles;
	\\
	{\bf Output:} $Q$ number of physical qubits used by factory.
	\begin{enumerate}
		\item $p_{\operatorname{tol}} \leftarrow 1/(3N)$
		\item $i \leftarrow 0$
		\item While $p_{\operatorname{tol}} < p_g$:
		\begin{enumerate}
			\item $i \leftarrow i+1$
			\item $d_i \leftarrow \min \{ d \in \N: 250 d (100 p_g)^{(d+1)/2} \ge p_{\operatorname{tol}} \}$
			\item $Q_i \leftarrow 15^{i-1}100 d_i(d_i-1)$
			\item $T_i \leftarrow 10 d_i$
			\item $p_{\operatorname{tol}} \leftarrow (p_{\operatorname{tol}}/36  )^{1/3}$
		\end{enumerate}
		\item layers $\leftarrow i$
		\item $S=\sum_{j=1}^{\mathrm{layers}} Q_j T_j$
		\item  Return $Q=N S / t_{\mathrm{algo}}$. 
	\end{enumerate}
}

We give a brief justification of these numbers for the Toffoli factory; the justification for the T state factory is essentially the same with a few steps removed.   The variable $p_{\operatorname{tol}}$ is initially set to the target error probability per Toffoli magic state.   We aim for $N p_{\operatorname{tol}} \leq 1/3$ so the total failure probability of the algorithm is less than a third. We work backwards from the last round of distillation to the first round, increasing $p_{\operatorname{tol}}$ until $p_{\operatorname{tol}}>p_g$.  Therefore we are assuming the initial magic states must have an error probability of $p_g$ or less, which is justified due to Ref.~\cite{li2015magic}.   The index $i$ starts at 1 for the last round of distillation that uses the Toffoli protocol, and we increment $i$ as we go to lower rounds of distillation. A key concept here is that of balanced investment~\cite{ogorman17}.  That is, on the $i^{\mathrm{th}}$ round of distillation we use a surface code of distance $d_i$ that suffices to reduce Clifford gate noise below the target error probability $p_{\operatorname{tol}}$. 

In steps (3)-(5) we calculate the resource costs in the Toffoli routine~\cite{eastin13,jones13,ogorman17}.   The noise per logical Clifford gate is suppressed to $\sim d(100 p_g)^{(d+1)/2}$ and there are 99 possible locations for a logical gate error (this will become clear below). Given the surface code distances, we can calculate the quantity of physical qubits needed.  Each Toffoli routine uses 11 logical qubits when ancillae are included (see Fig.~9 of Ref.~\cite{ogorman17}). For each logical qubit we need $2 d_i(d_i -1)$ physical qubits to realise the code and a further $2 d_i(d_i -1)$ qubits for syndrome extraction.  Therefore we require $44 d_i(d_i -1)$ physical qubits as in step (4).  The protocol is depth 9 in logical gates, with each logical gate requiring $d_i$ surface code cycles, giving step (5). For a depth 9 circuit with 11 logical qubits, the number of possible error locations is upper bounded by the value 99 that we used earlier.  The Toffoli routine takes $T$ magic states of error probability $\epsilon$ and outputs a Toffoli state with error probability $27\epsilon^2 + O(\epsilon^3)$, which for $\epsilon<10^{-3}$  is strictly upper bounded by $28 \epsilon^2$.  Inverting this relationship, we obtain the $p_{\mathrm{tol}}$ update in step 6. There is some finite failure probability but for $p_g \leq 10^{-4}$ this will not affect the order of magnitude of the calculation.  

Next, we iterate through a suitable number of rounds of the $15 \rightarrow 1$ Bravyi-Kitaev protocol~\cite{bravyi05a}.    The noise per logical Clifford gate is suppressed to $\sim d_i(100 p_g)^{(d_i+1)/2}$ and there are 250 possible locations for a logical gate error (this will become clear below). Given the surface code distances, we can calculate the quantity of physical qubits needed.    We use 25 logical qubits for the $15 \rightarrow 1$ protocol  (see Fig.~8 of Ref.~\cite{ogorman17}).   Therefore, $(2+2)*25 d_i(d_i -1) = 100 d_i(d_i -1)$ physical qubits are needed for each $15 \rightarrow 1$ protocol.  On the distillation round with index $i$, we need $(8*15^{i-2})$ copies of the $15 \rightarrow 1$ protocol.  The product of these numbers gives $Q_i$ in line 7(c).  The $15 \rightarrow 1$  protocol can be executed in depth 10 logical gates each requiring $d_i$ surface code cycles, giving a total $T_i$ shown in step 7(d).  Since there are 25 logical qubits in a depth 10 circuit, this gives the 250 potential error locations asserted earlier.   We must update $p_{\mathrm{tol}}$.  The $15 \rightarrow 1$ routine takes $T$ magic states of error probability $\epsilon$ and outputs a T state with error probability $35\epsilon^3 + O(\epsilon^4)$, which for $\epsilon<10^{-3}$ is strictly upper bounded by $36 \epsilon^3$.  Inverting this relationship, we obtain the $p_{\mathrm{tol}}$ update in step 7(e).  Again, the failure probabilities are negligible in the regimes considered here.

Finally, we combine the physical qubit cost and time cost into a single total space-time cost $S$ per Toffoli state, shown in step (9).  To deliver $N$ of these states within time $t_{\mathrm{algo}}$ requires a factory with a total number of qubits given by step (10).  The constant $t_{\mathrm{algo}}$ must be input in units of surface code cycles.
 
 We offer some remarks on possible additional savings. There is potential to cut these numbers in half by using the rotated surface code~\cite{horsman2012surface} but it is currently unknown whether the error suppression still obeys $\sim d(100 p_g)^{(d+1)/2}$ so we instead opt for a conservative, well-supported estimate.   There are several additional protocols for T state distillation (including Refs.~\cite{bravyi2012magic,Meier13,campbell2018magic}) but optimising over all these protocols is much more involved and the above estimate will give a similar order of magnitude result.    It is known that Toffoli states can be distilled using 6 T-states (asymptotically) rather than 8 T-states (see Refs.~\cite{campbell2017unified,campbell2017unifying}) but we do not know the ancilla or time cost of implementing this protocol.  It is often hoped that significant savings could be made by circumventing the need for magic state factories altogether, but the resource trade-offs are subtle; see Ref.~\cite{campbell2016steep} for a review. 
 
\bibliographystyle{plainnat}
\bibliography{aquacsp_doi}

\onecolumn

\begin{figure}[h!]
\begin{center}
\includegraphics{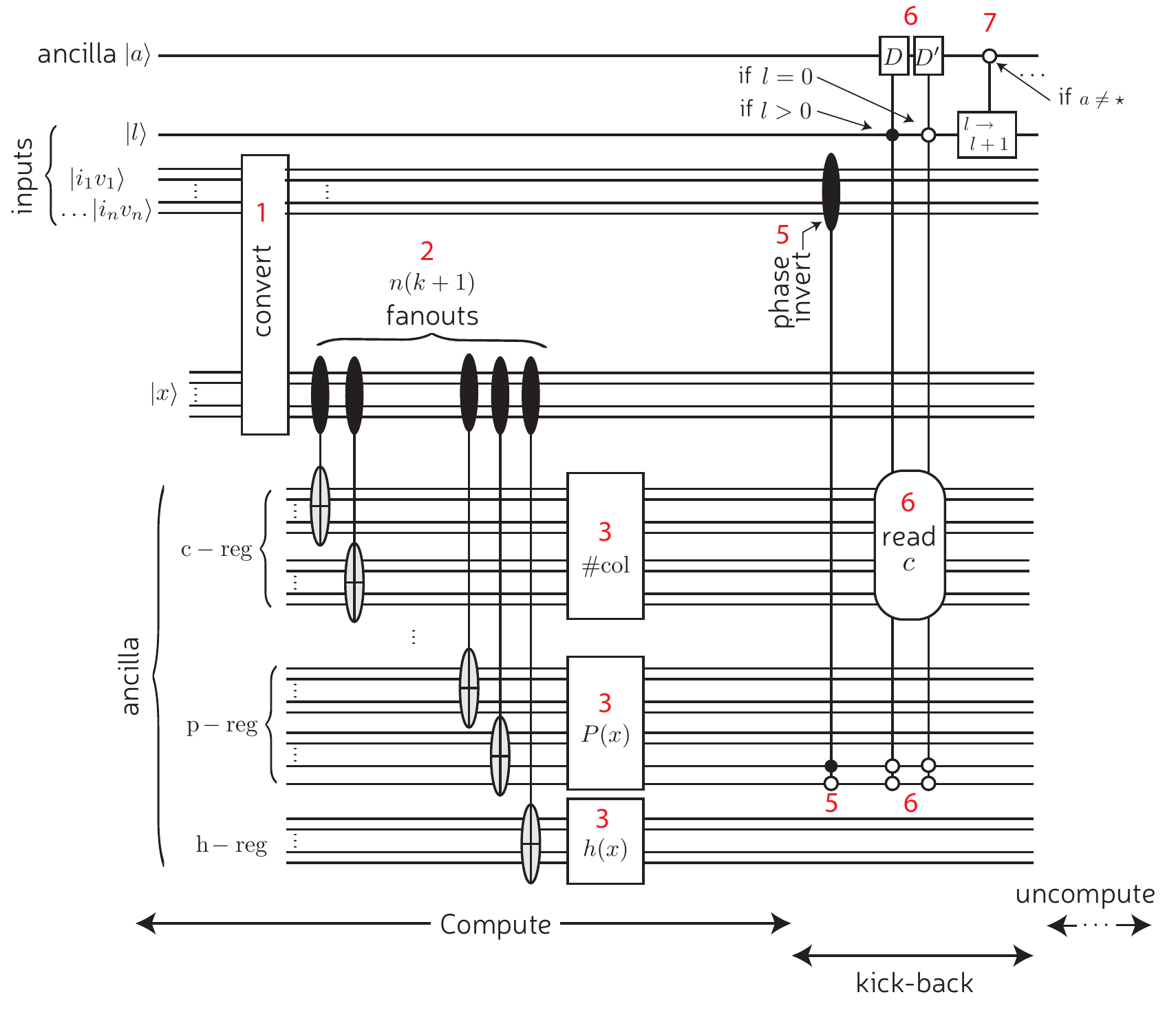}
\end{center}
\caption{Quantum circuit diagram for part of Algorithm \ref{alg:optbt} implementing $R_A$ in the special case where the level $\ell$ is assumed to be even. For illustrative purposes only as some aspects may be implemented differently to the diagram. The evaluation of $h(x)$ is used in the uncomputation step (not shown).}
\label{fig:ra}
\end{figure}

\end{document}